\pdfoutput=1
\documentclass[reprint,superscriptaddress]{revtex4-1}

\usepackage[T1]{fontenc} 
\usepackage[utf8]{inputenc}
\DeclareUnicodeCharacter{2212}{-}

\usepackage[ngerman]{datetime}
\newdateformat{myformat}{\THEDAY{. }\monthnamengerman[\THEMONTH] \THEYEAR}
\usepackage{parskip} 
\usepackage{svg}
\usepackage{import}
\usepackage{color}
\usepackage{float}
\usepackage{subfig}

\usepackage{tikz}
\usepackage{pgfplots}
\usepgfplotslibrary{external} 
 \tikzset{external/system call ={lualatex
 \tikzexternalcheckshellescape
 -halt-on-error
 -interaction=batchmode
 -jobname "\image" "\textsource"}}
\usepackage{amsmath} 
\usepackage{amssymb} %
\usepackage{wasysym} %
\usepackage{marvosym}
\usepackage{mathtools}
\usepackage{upgreek}
\usepackage{xcolor}
\usepackage{paralist}
\usepackage{enumitem} 
\usepackage{xcolor,hyperref}
\hypersetup{ 
	colorlinks=true,       
    linkcolor=blue,          
    citecolor=blue,        
    filecolor=magenta,      
    urlcolor=cyan           
    }
\newcommand{\subR}[1]{
  \protect\begin{NoHyper}\protect\subref{#1}\protect\end{NoHyper}
}

\newcommand{\mathsym}[1]{{}}

\newcommand{\LP}{\Delta}

\newcommand{\vc}[1]{\boldsymbol{#1}}

\newcommand{\partd}[1]{\partial_{ #1}}
\newcommand{\bra}[1]{\left( #1 \right)}
\newcommand{\brc}[1]{\left[ #1 \right]}
\newcommand{\mean}[1]{\left< #1 \right>}

\newcommand{\g}[1]{\begin{gather}      #1     \end{gather}}

\newcommand{\alg}[1]{\begin{align}      #1     \end{align}}

\DeclareMathOperator{\inc}{inc}
\DeclareMathOperator{\sgn}{sgn}
\begin{document}%
\title{How to Pare a Pair: Topology Control and Pruning in Intertwined Complex Networks.}

%
\author{Felix Kramer}
\affiliation{International Max Planck Research School for Cell, Developmental and Systems Biology,  Dresden 01307, Germany}
\affiliation{Max Planck Institute for Molecular Cell Biology and Genetics (MPI-CBG),  Dresden 01307, Germany}
\affiliation{Center for Systems Biology Dresden (CSBD),  Dresden 01307, Germany}
\author{Carl D. Modes} 
\affiliation{Max Planck Institute for Molecular Cell Biology and Genetics (MPI-CBG),  Dresden 01307, Germany}
\affiliation{Center for Systems Biology Dresden (CSBD),  Dresden 01307, Germany}
\affiliation{Cluster of Excellence, Physics of Life, TU  Dresden, Dresden 01307, Germany}

\date{\today}



\begin{abstract}
Recent work on self-organized remodeling of vasculature in slime-mold, leaf venation systems and vessel systems in vertebrates has put forward a plethora of potential adaptation mechanisms. All these share the underlying hypothesis of a flow-driven machinery, meant to alter rudimentary vessel networks in order to optimize the system's dissipation, flow uniformity, or more, with different versions of constraints. Nevertheless, the influence of environmental factors on the long-term adaptation dynamics as well as the networks structure and function have not been fully understood. Therefore, interwoven capillary systems such as found in the liver, kidney and pancreas, present a novel challenge and key opportunity regarding the field of coupled distribution networks. We here present an advanced version of the discrete Hu--Cai model, coupling two spatial networks in 3D.  We show that spatial coupling of two flow-adapting networks can control the onset of topological complexity in concert with short-term flow fluctuations. We find that both fluctuation-induced and spatial coupling induced topology transitions undergo curve collapse obeying simple functional rescaling. Further, our approach results in an alternative form of Murray's law, which incorporates local vessel interactions and flow interactions. This geometric law allows for the estimation of the model parameters in ideal Kirchhoff networks and respective experimentally acquired network skeletons.
\end{abstract}

\maketitle

\section{Introduction}
Many recent studies on biological transportation networks have been focused on the hypothesis that vasculature is remodeled according to the flow-induced stress sensed by the cells making up the tissue \cite{LENOBLE:2005fu}. This self-organized process optimizes the structures for the task at hand , e.g.\ distributing oxygen and nutrients, getting rid of waste,  carrying local secretion. The actual tissue response is dependent on the time-scales probed. On the one hand, short-term changes usually concern rapid vessel diameter changes in response to pressure fluctuations or medication. On the other hand, long-term effects e.g.\ due to metabolic changes may manifest in permanent diameter changes \cite{2012PLoSO...745444H}, usually leaving the vessel structure with a trade-off between efficiency and redundancy \cite{Ronellenfitsch:2019fe}.  \\
Particular focus has been directed to the long-term remodeling of the capillary plexus and other rudimentary transport systems in the early developmental stages of organisms, i.e.\ by studying complex signaling cascades involving growth factors like VEGF in vascular systems of mammals \cite{1997Natur.386..671R} or auxin in plants \cite{Dimitrov:2006gw}. Yet, the onset of refinement seems to be correlated with mechanical stresses (such as shear flow) as has been shown in a variety of model organisms from chicken embryo \cite{LENOBLE:2005fu,Nguyen:2006kr} and zebrafish \cite{Lenard:2015de} to leaves \cite{RothNebelsick:2001dh} and slime mold \cite{Tero:2007hh}.\\
Early theoretical approaches by Murray \cite{Murray:1926tj,1926PNAS...12..299M} posited that diameter adaptation would minimize the overall power dissipation of the system. Following this ansatz of network optimization, many recent models are using global optimization schemes on expanded vessel networks. These models captured the phenomenon of link pruning involving random damage, flow fluctuations or rescaled volume costs and have been able to account for the trade-off of shunting and redundancies \cite{Bohn:2007fi, Katifori:2010kt, Corson:2010ee}. Further advances have been made in empirical studies of local vessel dynamics, e.g.\ blood vessel systems \cite{Pries:1998tn, Pries:2001vg, Secomb:2013hm}. Local adaptation dynamics were also effectively derived by minimizing various effective network costs via gradient descent methods \cite{Hu:2013io,Chang:2019ip}. It has further been shown that the outcomes of locally adapting networks are robust against variations of the initial topological structure \cite{Grawer:2015bh} and that plexus growth and correlated flow fluctuations can provide elaborate hierarchies \cite{Ronellenfitsch:2016hh, Ronellenfitsch:2019fe}. Many of these effects may also be seen in continuous adaptation models in porous media  \cite{Haskovec:2015ei,Haskovec:2016go}. It is interesting and important to note here that these adaptation mechanisms may leave certain fingerprints, e.g.\ in the form of allometric \cite{West:1997cg} and geometric laws \cite{Sherman:1981ux}.\\
These studies typically involve volume or metabolic constraints applied to abstract Kirchhoff networks. Yet they disregard the key characteristic common to all fluid transport systems: spatial embedding, which matters especially in the case of capillary systems as these directly interact via transport of metabolites with the surrounding tissue. These systems have to maintain a robust structure while being embedded in a possibly stiff tissue environment potentially perturbing the shear stress driven adaptation mechanism.\\
We here focus on the development and function of multicomponent flow networks, which influence each other based on their spatial architecture. Biologically speaking, these systems often consist primarily of blood vessels and a secondary entangled, interacting system as found for example in liver lobule \cite{Lautt:2007be, Boyer:2013hj, Meyer:2017fh, MoralesNavarrete:2015kk}, the kidney's nephrons \cite{Shah:2004jf,Serluca:2002vj}, the pancreas \cite{Magenheim:2011bz,Villasenor:2012js,Azizoglu:2016ip} or the lymphatic system \cite{PlanasPaz:2013fj}. Additionally, we intend to include the phenomenon of one flow network being `caged' by another complementary structure, e.g.\ capillaries embedded in bone marrow \cite{Sivaraj:2016ip}.\\ In this work we study the adaptation of two coupled spatial networks according to an advanced version of the discrete Hu--Cai model \cite{Hu:2013io} including Corson fluctuations \cite{Corson:2010ee}. Each network is subject to flow driven and volume-constrained optimization on its own. Meanwhile we introduce the networks' interaction in the form of a mutual repulsion or attraction of vessel surfaces. Repulsion will prevent them from touching directly by their otherwise flow driven radius expansion, introducing a competition of the two networks for the space they are embedded in. Alternatively, vessel surfaces could be be attracted towards each other, presenting a positive feedback towards maintaining intertwined structures. In combination with fluctuation induced redundancy, we find mutual repulsion to greatly reduce the networks relative loop density even when strong fluctuations are present. On the other hand we observe the emergence of a new sharp transition towards a loopy state when there is attraction between the two networks.\\
Unfortunately few metrics provide the means to fit or estimate the applied parameters of adaptation models for real systems, even though time-lapse experiments \cite{Lenard:2015de}, counting pruning events and topology analysis on pruned structures \cite{Ronellenfitsch:2019fe,Modes:2016df,Papadopoulos:2018em} allow for qualitative insights into the mechanism at hand for certain model organisms. Yet there has been no proposal to our knowledge to quantitatively acquire or fit the model parameters from real, pruned network structures. In particular, interwoven systems present a special challenge as typical experimental setups for a full 3D reconstruction involve invasive measures, i.e.\ sacrificing the specimen and preventing any long-term vessel observation. To tackle this problem we generalize an important scaling law, which has been discussed again recently in this context \cite{Akita:2016fp}: Murray's Law. 
Our generalization enables us to reconstruct the model parameters with high fidelity for Kirchhoff networks solely from a given graph topology and it's edge radii distribution (assuming the pruning process reached a stationary state). We find order of magnitude estimate for these parameters in experimentally acquired data sets of an interwoven system: sinusoids and canaliculi in the mouse's liver acinus.\\
We begin our study in section \ref{sec:theory} with a brief reminder of the hydrodynamical and network theoretical background on which we operate. Next we set up our model framework and its crucial components in detail. In section \ref{sec:results} we present the numerical evaluation of our model, in particular illuminating the cases of repulsive and attractive coupling between networks. In section \ref{sec:estimation} we derive and test our generalized geometric laws on ideal Kirchhoff networks and on datasets of vessel networks provided by our collaborators. We then go on to discuss the implications and limits of our model framework in the concluding section \ref{sec:discussion}.
\section{Theoretical Framework}\label{sec:theory}
The following subsections are intended to provide the reader with the necessary background to proceed to the complex adaptation dynamics on ramified vascular networks that follow. Readers familiar with the general formalities should feel free to skip ahead to section \ref{sec: intertwined_model} where we discuss our general set up.\\ First we introduce the framework of Kirchhoff networks as these provide us with the mathematical tools needed to describe complex flow landscapes. Afterwards, we reintroduce the cost-function ansatz, the associated metabolic costs, and the chosen method of optimization which will render towards adaptation dynamics. Next, we discuss the intended hydrodynamic regime and the geometry of the intertwined system. We then extend our established framework by including fluctuation of the flow landscape as an essential tool to generate robust distribution networks. Finally, in the last subsection we introduce the relevant order parameters and metrics.

\subsection{Fundamentals of linear networks}
We model the  biological vessel networks of interest as a composition of $m$ edges and $n$ vertices (branching points). Each edge carries a flux $f$ such that at any vertex the sum of all currents equal a nodal function $s$, 
\g{
	\sum_{e\in  \inc(v)} f_e  =s_v \label{eq:current_law}
}
where $\inc(n)$ indicates the set of edges incident to vertex $v$. We refer to $s$ as sink or source when $s$ is non-zero. Equation \eqref{eq:current_law} is Kirchhoff's current law, which represents mass conversation at every vertex. Further, in linear flow networks one may formulate the flux as a linear function, Ohm's law, as:
\g{
	f_e=  c_e \LP p_e \label{eq:ohm_law}
}
where $c_e$ is the conductivity of an edge and $\LP p_e$ its respective potential gradient. One may thus characterize the flux in every vessel as a direct response to a gradient of potential energy, concentration, temperature etc. and have it scale linearly with the conductivity which incorporates the geometry and physical nature of the transport problem. The equation systems formulated in \eqref{eq:current_law} and \eqref{eq:ohm_law} may be bundled in vectorial notation as:
\g{
	\vc{B}\cdot \vc{f}=\vc{s} \label{eq:current_law_vec}\\
	 \vc{f}=\vc{C}\cdot\LP\vc{p}\label{eq:ohm_law_vec}
}
Here $\vc{B}$ designates the incidence matrix and $\vc{C}$ is a diagonal matrix with $c_e$ on the diagonal. Combining equations \eqref{eq:current_law_vec} and \eqref{eq:ohm_law_vec} one finds the transformation between the sinks/sources and the potentials as:
\g{
	\vc{B}\cdot \vc{C}\cdot\LP\vc{p}=\vc{s}\label{eq:trafo}
}
Unfortunately, this equation system is under-determined and accordingly seems to lack a unique solution for $\LP\vc{p}$. 
Interestingly enough one may find a unique solution to the problem \eqref{eq:trafo} by applying the Thomson principle \cite{KELLY:1991hd,Grady:2010vv}. Following the Thomson principle, one considers the system to be characterized by a cost function, i.e.\ the energy dissipation defined as $\sum_e f_e^2/\tilde{c}_e$, with positive coefficients $\tilde{c}_e \geq 0$. Further one may use this cost to formulate an optimization problem with Lagrange multipliers $p_v$ and the boundary conditions \eqref{eq:current_law} such that:
	\g{
	\Gamma=\sum_e f_e^2/\tilde{c}_e +\sum_v p_v\bra{s_v - \sum_{e' \in  \inc(v)} f_{e'} }\label{eq:lagrange}
	}
The aim is to find the set of flows $f_e$ which minimize the system's cost, $\Gamma$, with respect to the constraints given by the Kirchhoff current law. Doing so one will end up naturally with Ohm's law, with the conductivities $c_e=\tilde{c}_e$ for the coefficients  and the Lagrange parameters $p_v$ representing the nodal potentials. These $p_v$ define the potential gradients as $\LP p_e= \bra{ p_{\omega(e)}-p_{\alpha(e)}}$, where $\alpha(e), \omega(e)$ designate the initial and final vertex of any edge $e$. This cost function ansatz enables one to find a unique solution for the potential differences $\LP p$ in equation \eqref{eq:trafo} as:
\g{
	\LP\vc{p}=\vc{C}^{-1/2}\brc{\vc{B}\cdot\vc{C}^{1/2}}^{\dagger}\vc{s} \label{eq:pressure_solution
}}
where $\brc{\cdot}^{\dagger}$ designates the generalized inverse \cite{1955PCPS...51..406P}. This solution represents the optimal potential landscape, which minimizes the overall power dissipation for a given landscape of conductivities and sinks  \cite{BenIsrael:1963kd,1956PCPS...52...17P}.
Note that this formalism may be applied to any stationary transport process following the Thomson principle as well as random walks of particles on a lattice. This class of systems is often referred to as lumped systems or Kirchhoff networks, in analogy to simple electric circuits \cite{TN_libero_mab214116274}. 

\subsection{Cost function ansatz and optimization of biological networks } \label{sec:cost_ansatz}
The concept of characterizing a transport network by a cost may readily be transferred to dynamic biological systems. The cells which are forming the walls of vascular networks for example, are able to respond and adapt to a given set of stimuli such as shear stress or hydrostatic pressure. This enables such systems to continuously change their own topology and edge conductivities in order to reach final refined structures. To capture this behavior one may formulate a cost for a vessel systems as proposed in \cite{Bohn:2007fi}:
\g{
	\Gamma=\sum_e \frac{f_e^2}{c_e} +a c_e^{\gamma} \label{eq:cost_ansatz}
}
where the first term is the power dissipation as before and the second a metabolic cost term $c_e^{\gamma}$, with proportionality factor $a$. This second term encapsulates the notion that a biological organism is constrained by the metabolic costs to deploy and sustain a vessel of a certain conductivity. The exponent $\gamma$ represents a degree of freedom to vary the relative importance of vessels of low or high conductivity. \\The minimization of the function \eqref{eq:cost_ansatz} is performed by finding the set of conductivities $c_e$ which minimizes \eqref{eq:cost_ansatz} for a given boundary condition $\vc{s}$. Following the ansatz in \cite{Hu:2013io} we may formulate our minimization in the form of temporal adaptation rules for each vessel, where each element reacts to a local stimulus, instead of a single global optimization procedure. To derive a local adaption dynamic we perform a gradient descent approach. This means we consider the temporal derivative $\Gamma$ as:
\g{
	\frac{d\Gamma}{dt}=\sum_j \bigtriangledown_{c_e} \Gamma\cdot \partd{t}c_e\label{eq:gradient_descent1}}
We want to ensure $\frac{d\Gamma}{dt} \leq 0$ and therefore that $\Gamma$ converges towards a local minimum. To do so we may formulate the dynamical equations for $c_e$ as the negative gradient of $\Gamma$:
\g{
 \partd{t}c_e\propto - \bigtriangledown_{c_e} \Gamma= \bra{ \frac{f_e^2}{c_e^{2}} - a\gamma  c_e^{\gamma-1}}\label{eq:gradient_descent2}
}
where we used the definition of $\Gamma$ in equation \eqref{eq:cost_ansatz}. The dynamics in equation \eqref{eq:gradient_descent2} allow for a continuous local adaptation of the vessel's state by consideration of its local flux, current conductivity and metabolic parameters $a, \gamma$. We extend this approach for interacting multilayer networks in a linear manner by adding up the metabolic cost of the individual systems involved and adding respective interaction terms. In our particular case we do so for two flow networks with:
\g{
	\Gamma= \Gamma_{12}+\Gamma_1+\Gamma_2 \label{eq:cost_function_joined}
}
where $\Gamma_1$, $\Gamma_2$ are given for each network by equation \eqref{eq:cost_ansatz}. The interaction term $\Gamma_{12}$ incorporates the geometrical nature of the system and encapsulates either a competition or symbiosis of the vessels of the two systems on a local basis as well. The exact nature of this interaction term as well as of the metabolic costs will be discussed in the next section in further detail, where we derive the dynamical systems in accordance with equation \eqref{eq:gradient_descent2}. 
\subsection{Modeling intertwined vessel systems}\label{sec: intertwined_model}
In this section we connect the results of the previous sections to the relevant hydrodynamics and discuss the nature of the exact biological flows problem. The direct applicability of equations \eqref{eq:current_law_vec}, \eqref{eq:ohm_law_vec} to biological systems becomes clear when considering the Hagen--Poiseuille law \cite{Landau:1959vq} which describes the volume flow rate $f$ of a fluid of viscosity $\eta$ at low Reynold numbers through a tube of length $l$ and radius $r$ as:
\g{
	f=\bra{\frac{\pi r^4}{8\eta l}}\Delta p \label{eq:ohm_poiseuille}
}
This provides us with a conductivity $c=\frac{\pi r^4}{8\eta l}$ by direct comparison with Ohm's law, with a fourth order dependency of the radius $r$. We will here focus on radial adaptation and consider the special case of $l_i=\text{const.}$ for all vessels in either network. Hence, using equations \eqref{eq:cost_ansatz}, \eqref{eq:ohm_poiseuille} we may rewrite the cost ansatz in \eqref{eq:cost_function_joined} as: 
\alg{
	\Gamma = & \Gamma_{12} + \bra{\frac{\pi}{8\eta_1 l_1}}\brc{\sum_{e_1} r_{e_1}^4\LP p_{e_1}^2 +a_1 r_{e_1}^{4\gamma_{1}}}\nonumber\\ & +\bra{\frac{\pi}{8\eta_2 l_2}}\brc{\sum_{e_2} r_{e_2}^4\LP p_{e_2}^2 +a_2 r_{e_2}^{4\gamma_{2}}}\label{eq:cost_function_joined_radial}
} 
From here on we use the indices $i\in\lbrace1,2\rbrace$ for the two networks. Further, we only consider the specific case $\gamma_{i}=0.5$, which relates the metabolic cost directly towards volume for each vessel. Performing a minimization of the cost \eqref{eq:cost_function_joined_radial} at this point, without considering any interactions $\Gamma_{12}$, would result in each network to independently become dissipation minimized, constrained by its overall volume. We thus turn our attention to the interaction term $\Gamma_{12}$.
\begin{figure}[h]
		 \def\svgscale{0.13}
			\subfloat[]{\includegraphics[scale=0.15]{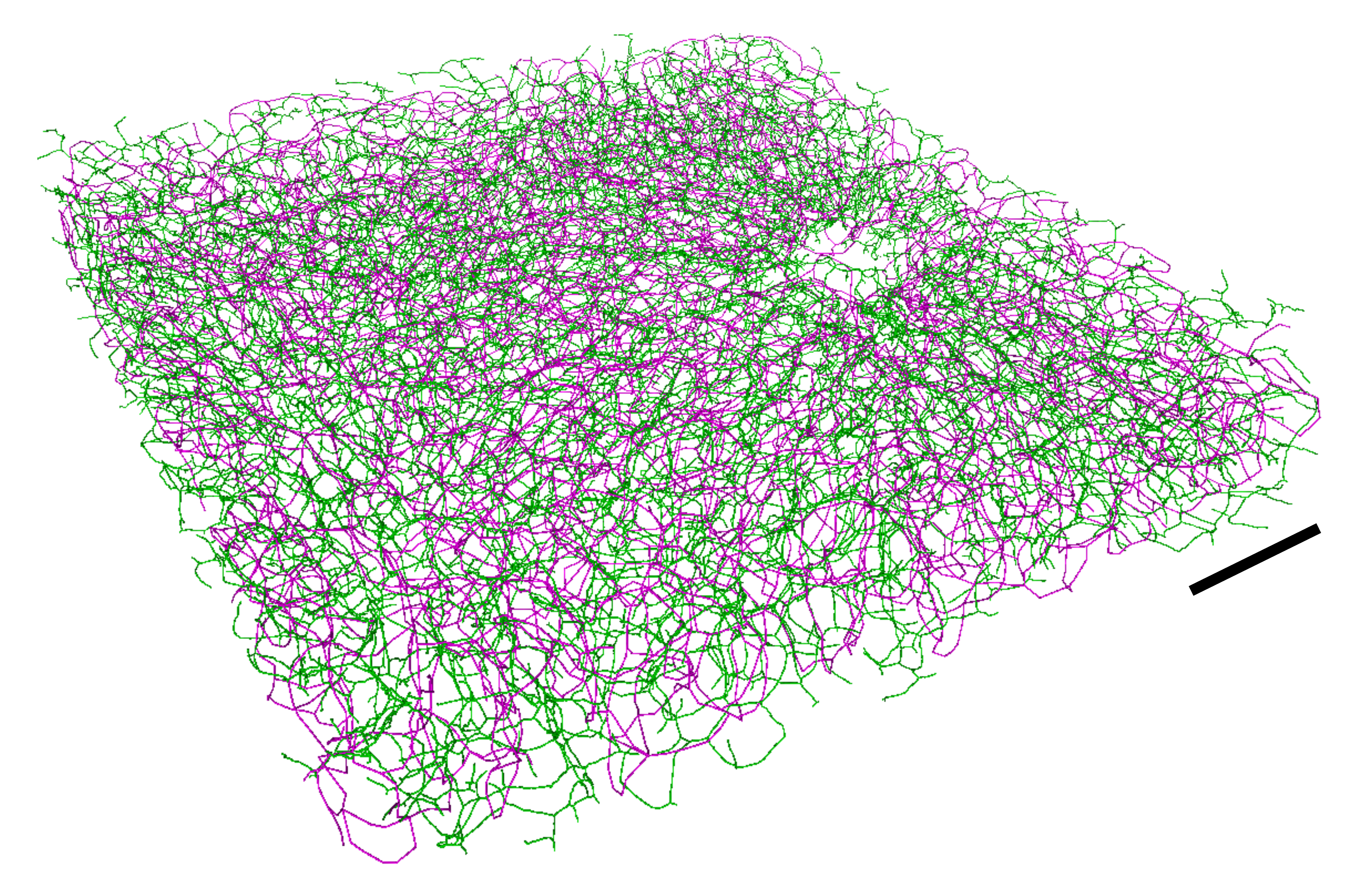}\label{fig:real_network}}\\
		 \def\svgscale{0.15}
			\subfloat[]{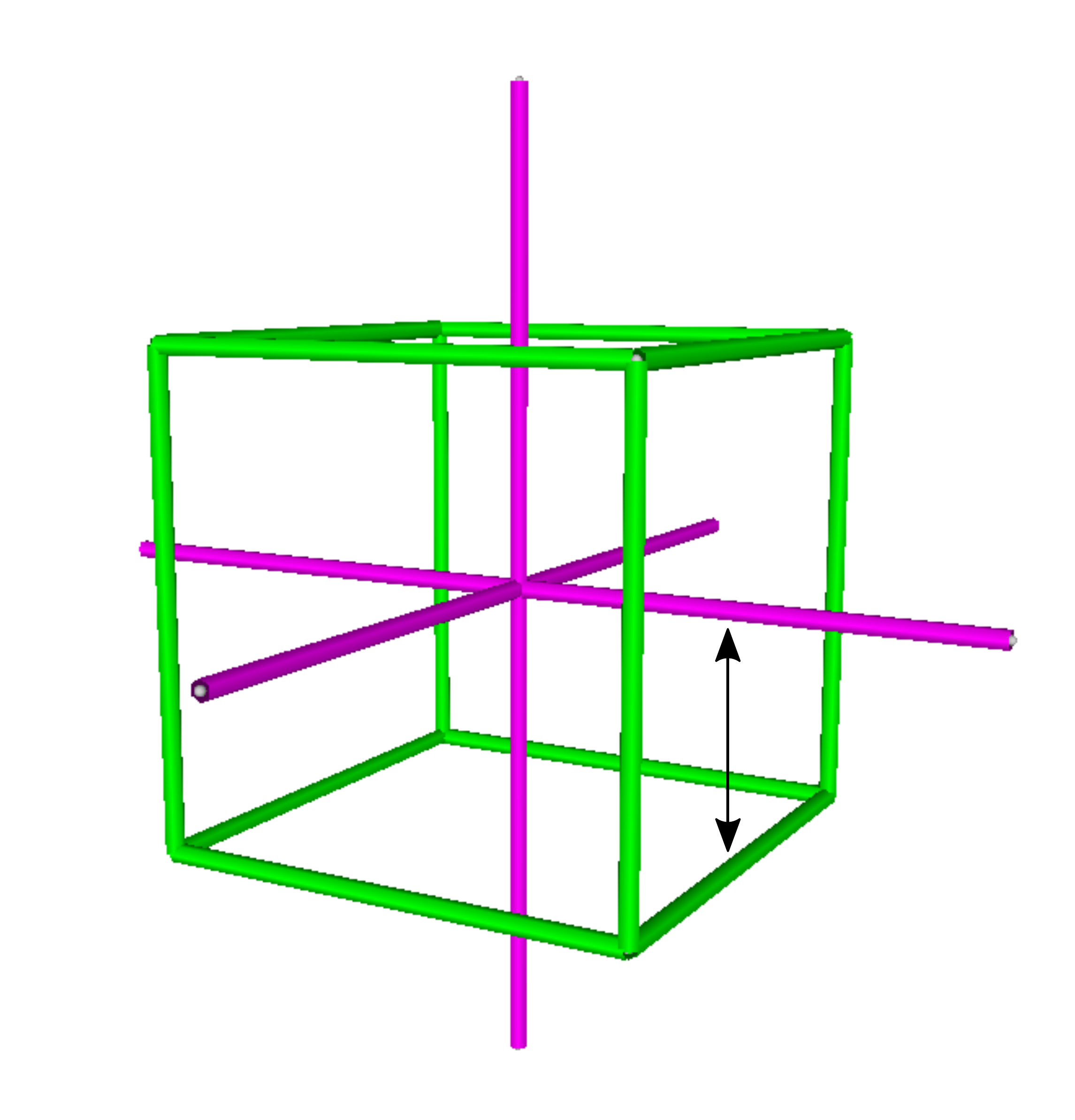\label{fig:model_a}}
			\subfloat[]{\includegraphics[scale=0.11]{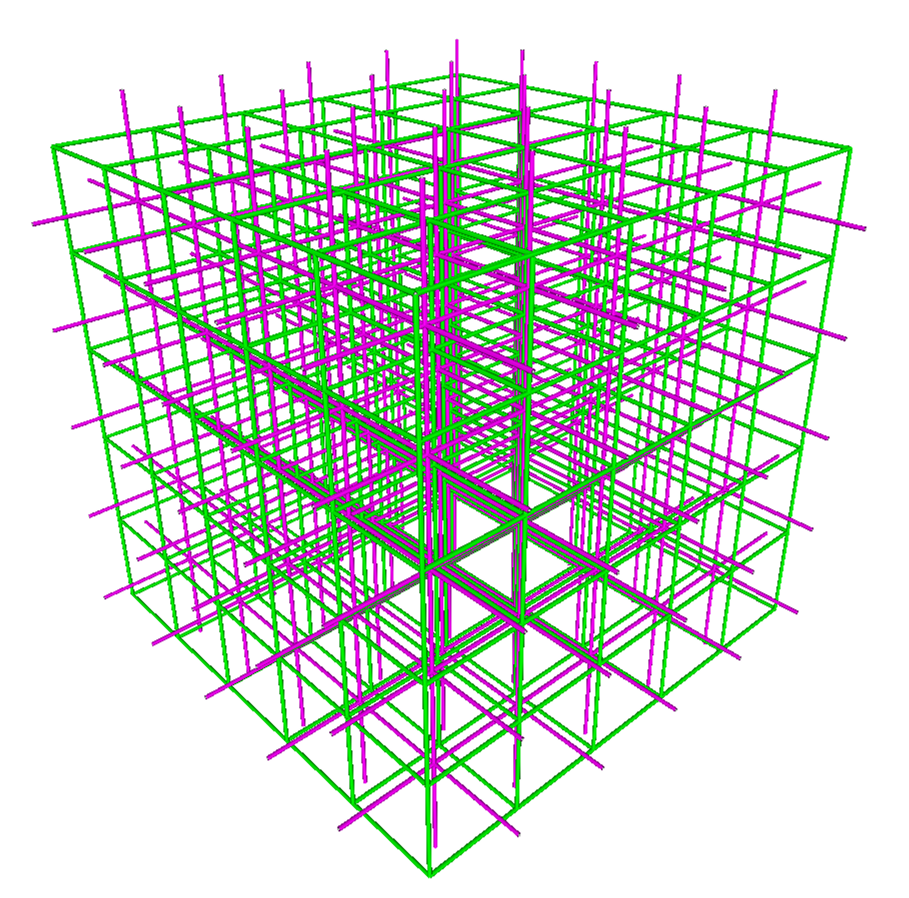}\label{fig:model_b}}
			\caption{\subR{fig:real_network}~Biological intertwined structures: segment of bile canaliculi (green) and sinusoids (magenta) in the mouse's liver acinus (presented as network skeletons, see also \cite{MoralesNavarrete:2015kk} )), scale bar is $200\ \mu m$. Modeling two interacting networks and 3D network skeletons of triply-periodic minimal surfaces for initial plexi: \subR{fig:model_a}~(Cubic) unit cell and indication of spatial constraint for tube surfaces. \subR{fig:model_b} ~Complementary cubic lattices. 
			 }
			\label{fig:entangled_model}
		\end{figure} 
In order to model interacting networks such as those found in the liver lobule (see Figure~\ref{fig:real_network}) we define a multilayer network consisting of two intertwined, yet spatially separate objects each consisting of edges $e_i$, with designated vessel radii $r_{e_i}$ see Figure~\ref{fig:model_a} and \ref{fig:model_b}. Each edge in either network is affiliated with the set of closest adjacent edges of the other respective network. As all theses edges are simply tubes in our model, we define the distance between affiliated tube surfaces to be,
\g{
		\LP r_{e_1e_2}=L-\bra{r_{e_1}+r_{e_2}}
}
where $L$ is the initial distance of the abstract network skeletons (equal to distance in case of simultaneously vanishing radii). To model a system of blood vessels entangled with a secondary, secreting vessel network, we postulate that the respective tube surfaces must not fuse or touch directly, i.e.\ $\LP r_{e_1e_2} \geq 0$. Subsequently, we construct the interaction term $\Gamma_{12}$ for the combined system as a power of the relative distance $\LP r_{e_1e_2}$:
\alg{
\Gamma_{12} & =\frac{b}{2}\sum_{e_1e_2}F_{e_1e_2}\LP r_{e_1e_2}^{\varepsilon}\\ \text{ with } F_{e_1e_2} & =\begin{cases}
1 & \text{if edges }e_1 \text{ and }e_2 \text{ affiliated}\\
0& \text{else}
\end{cases}
}
with positive coefficient $b \geq 0$ and exponent $\varepsilon \in \mathbb{R}$ allowing us to switch between a repulsive or attractive behavior of the interaction, see section \ref{sec:Adaptation}, resembling either the competition for space or a mechanism to increase mutual contact. We have therefore arrived at the total cost function for the system:
\alg{
	\Gamma = &\frac{b}{2}\sum_{e_1e_2}F_{e_1e_2}\LP r_{e_1e_2}^{\varepsilon} \nonumber\\ & + \bra{\frac{\pi}{8\eta_1 l_1}}\brc{\sum_{e_1} r_{e_1}^4\LP p_{e_1}^2 +a_1 r_{e_1}^{2}}\nonumber\\ & +\bra{\frac{\pi}{8\eta_2 l_2}}\brc{\sum_{e_2} r_{e_2}^4\LP p_{e_2}^2 +a_2 r_{e_2}^{2}} \label{eq:cost_function_joined_radial2}
} 
which we may now use to derive the dynamical system via gradient descent.
\subsection{Adaptation dynamics of intertwined vessel systems}\label{sec:Adaptation}
\begin{figure}[h]
\includegraphics[scale=1.]{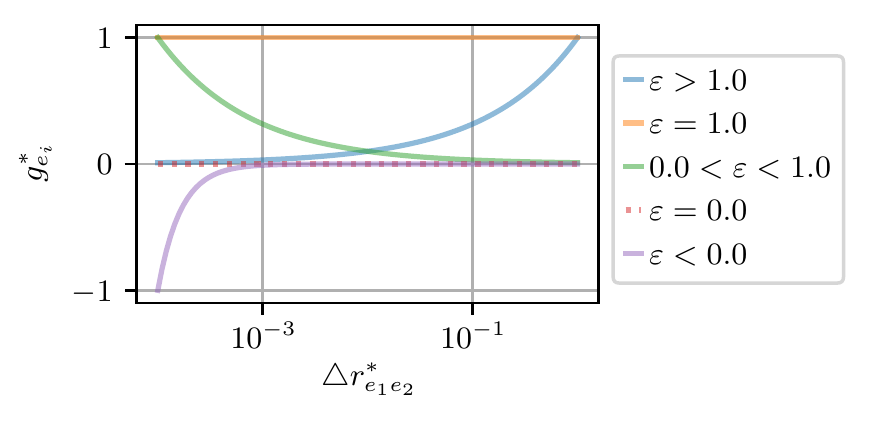}
\caption{The effective behavior of the coupling term $g_{e_{i}}^*$ changes with variation of the exponent $\varepsilon$ . Function values of $g_{e_{i}}^*$ are presented here dependent on the distance of a single arbitrary vessel pairing, curves normalized for comparison.
 }\label{fig:power_law}
 \end{figure}

In this section we discuss in detail the dynamical systems we intend to construct via the gradient descent approach on the basis of the cost function shown in equation \eqref{eq:cost_function_joined_radial2}. Calculating the gradient we acquire the equations of motion for each network as:
\alg{
 \partd{t}r_{e_{1}}  \propto &\brc{\LP p^2_{e_{1}} r_{e_{1}} ^2-\frac{a_1}{2}}r_{e_{1}} \nonumber\\ &+\frac{b\eta_1 l_1\varepsilon}{\pi }\sum_{ e_{2}}F_{e_1e_2}\LP r_{e_1e_2 }^{\varepsilon-1}\label{eq:gradient_1}\\
  \partd{t}r_{e_{2}}  \propto &\brc{\LP p^2_{e_{2}} r_{e_{2}} ^2-\frac{a_2}{2}}r_{e_{2}} \nonumber\\ &+\frac{b\eta_2 l_2\varepsilon}{\pi }\sum_{ e_{1}}F_{e_1e_2}\LP r_{e_1e_2 }^{\varepsilon-1}\label{eq:gradient_2}
}
The details of the derivation are given in the supplementary material \ref{apx:gradient_flow}. It may be noted here that the terms $\LP p_i^2 r_i^2$ correspond to the wall shear stress exerted, manifesting itself in a positive growth feedback. The negative terms $\frac{a_{i}}{2 }$ relate towards the metabolic cost. This term imposes effectively a volume penalty on the system as growing vessels generate an increased negative feedback. The interaction term reacts to the relative vessel distance $\LP r$ and imposes a feedback connected to the local neighborhood of each vessel which can be either positive (attractive coupling) or negative (repulsive coupling) depending on the choice of $\varepsilon$.\\
In order to perform a numerical evaluation of the resulting ODE system we define a unit system and non-dimensional parameters as follows: the radii and edge lengths in units of the grid distance $r_{e_{i}}=L r_{e_{i}}^*$ , $l_i=L l_i^*$, the nodal in and outflow $s_{v_{i}}=\mu_{i} s_{v_{i}}^*$, the conductivity $c_{e_{i}}=\eta_{i}^{-1}L^3 c_{e_{i}}^*$ and hence pressure $\LP p_{e_{i}}=\frac{\mu_{i}\eta_{i}}{L^3} \LP p_{e_{i}}^*$ and the networks' edge surface distance $\LP r_{{e_1e_2'}}^{\varepsilon-1}=L^{\varepsilon-1}\LP r_{e_1e_2}^{*\varepsilon-1}$. We define the time scale via the volume flow rates in the primal network as $t=\frac{L^3}{\mu_1}t^*$. Given positive proportionality constants $\chi_i \geq 0$ in the equations \eqref{eq:gradient_1}, \eqref{eq:gradient_2}, we define the effective temporal response parameters in either network as ${\lambda_0^i=\frac{\chi_{1}L^3}{\mu_1}\bra{\frac{\mu_i\eta_i}{L^2}}^2}$. Further we define the effective network couplings ${\lambda_1^i=\frac{b \vert\varepsilon\vert L^{3+\varepsilon}}{\pi \eta_i\mu_i^2}}$ and the effective volume penalties ${\lambda_2^i=\frac{a_i}{2}\bra{\frac{L^2}{\mu_i\eta_i}}^{2}}$. We introduce an effective coupling $g^*_{e_i}$ term as
\g{
	g_{e_{1}}^*=\sgn(\varepsilon)l_i^*\sum_{e_{2}}F_{e_1e_2}\LP r_{e_1e_2}^{*\varepsilon-1}\\
	g_{e_{2}}^*=\sgn(\varepsilon)l_i^*\sum_{e_{1}}F_{e_1e_2}\LP r_{e_1e_2}^{*\varepsilon-1}
	}
using the sign function $\sgn(\varepsilon)$. We thus arrive at the dimensionless form of the dynamical equations for each network:
\g{
\partd{t^*}r_{e_{i}}^*  = \lambda_0^i \bra{ \LP p_{e_{i}}^{*2}r_{e_{i}}^{*3} -\lambda_2^i r_{e_{i}}^*+ \lambda_1^i g_{e_{i}}^*} \label{eq:radius_scaled_0}
}
The coupling $g_{e_{i}}^*$ changes its qualitative behavior with variation of the exponent $\varepsilon$. We will consider the cases of an attractive coupling  $\varepsilon > 1$ and a repulsive coupling $\varepsilon < 0$, see Figure~\ref{fig:power_law}. We will further discuss the trivial case of an uncoupled system $\varepsilon=0$ (for which $\lambda_1=0$) for direct comparison with the Hu--Cai model. Note that $\varepsilon=1$ will result in a constant $g_{e_{i}}^*$ corresponding to a positive constant background stimuli, as proposed by \cite{Ronellenfitsch:2016hh}. The case $0<\varepsilon<1$ results in a diverging attraction for nearby vessels, while it basically drops out for $\LP r_{e_1e_2}\longrightarrow 1$. These numerically and heuristically unfavorable cases of $0<\varepsilon<1$ as well as the linear case $\varepsilon=1$ will be neglected hereafter.
\subsection{Incorporating flow fluctuations: Noisy, uncorrelated sink patterns}
Next we assume the adaptation of the vascular networks depends on an averaged potential landscape instead of instantaneous configurations, which are bound to occur in real systems due to short-term metabolic changes or vessel blocking/damage. In other words, we assume  a constant vessel radius between two adaptation events, while the flow rates change throughout the system due to changes in the sinks' magnitude, enabling us to substitute ${ \LP p_e^2\rightarrow \mean{\LP p_e^2} }$ in equation \eqref{eq:radius_scaled_0}. We thus implicitly assume a time-scale separation between the radii adaptation (long-time changes, not to be confused with short term contraction/dilation) and  changes of hydrostatic pressure. We define fluctuations in accordance to the Corson model \cite{Corson:2010ee}. Subsequently, we will only consider $\vc{s}$-configurations in which there exists one source-node (here $v_{\text{source}}=0$, $s_0=-\sum_{v>0}s_v$) and all other nodes are randomly initialized sinks with the following characteristics:
		\g{\mean{s_v}=\mu_v \text{ with }v>0\\ \mean{s_v s_w}=\rho_{vw}\sigma_v\sigma_w+\mu_v\mu_w \text{ with } v,w > 0 
		\label{eq:sink_condition}
		}
 We assume the fluctuations are uncorrelated and follow the same probability distribution. We set for the mean $\mu_v=\mu$ , standard deviation $\sigma_w=\sigma$, and correlation coefficient $\rho_{vw}=\delta_{vw}$. We may subsequently calculate the average squared pressure:
\g{
\mean{ \LP p_e^2 }  = \mu^2 \phi_e + \sigma^2 \delta\phi_e \label{eq:sq_pressure_solution}
}
where the function $\phi_e$ describes the squared pressure in the case of a constant source-sink landscape in the absence of any variance $\sigma^2$. Further, the function $\delta \phi_e$ describes the pressure perturbation caused by fluctuations with variation $\sigma^2$ in analogy to the Hu--Cai models heuristic fluctuation ansatz \cite{Hu:2013io}. For the full derivation of equation \eqref{eq:sq_pressure_solution} and the detailed computation of $\phi_e$, $\delta \phi_e$  see supplementary material \ref{apx:flow_fluctuation}. Using this approach we prevent shunting and the generation of spanning trees, which is caused by the typical `single source/multiple sinks' setting. Further, using this ansatz one also prevents accidentally partitioning the graph which can happen when realizing the sink-source configurations one by one \cite{Ronellenfitsch:2019fe}. \\We find this ansatz in particular fitting to model the liver lobule system, as sinusoids are fenestrated structures (meaning the vessel wall is perforated). Additionally, bile and water is  frequently secreted by hepatocytes (cells forming the bulk of the tissue and the basic metabolic unit in the liver) into bile canaliculi. On the other hand, one may argue that the fluid leak in the sinusoidal system is negligible in comparison to the overall throughput rate, and an additional distinguished sink would have to be placed at the opposing end of the plexus, extracting the majority of fluid. Here we neglect this factor as one major sink would merely generate one (or a small number of) distinguished large vessel(s), without any further impact on the topological complexity of the rest of the networks.\\
We incorporate these flow fluctuations with an effective fluctuation strength ${\lambda_3^i=\frac{\sigma_i^2}{\mu_i^2}}$ in equation \eqref{eq:radius_scaled_0} :
\g{
\partd{t^*}r_{e_i}^*  = \lambda_0^i\bra{ \bra{\phi_{e_i}^*+\lambda_3^i \delta\phi_{e_i}^*}r_{e_i}^{*3} -\lambda_2^i r_{e_i}^* + \lambda_1 g_{e_i}^*\label{eq:radius_scaled_3}}
 }
Thereby $\lambda_3^i$ scales the strength of pressure perturbations, which effectively impose an increase in the wall shear stress term in equation \eqref{eq:radius_scaled_3}.
\subsection{Order parameters for network remodeling}
In order to quantify the topological changes occurring in an adapting system we monitor the relative cycle density as an indicator of redundancy. Loosely, one may identify the number of cycles in a network in the following simple way: If we assume that the network's representing simple graph is one connected component of $n$ vertices and $m$ edges, then we only need $n-1$ edges to connect every vertex into a spanning tree without a single cycle, while each additional edge added from here on will form a cycle. Thus the total amount of such cycles in a network $z$, is the number of excess edges from the total number of edges:
\g{
	z= m - \bra{n-1} \label{eq:cycles}
}
Strictly speaking, $z$ is the amount of \textit{independent} cycles (also referred to as nullity) and may be calculated for any simple multicomponent graph \cite{Whitney:1932bg}. We use this metric in the following way: We solve the dynamical systems \eqref{eq:radius_scaled_3} until the networks reach a stationary state. The initial graph structure this process starts from is called a plexus and represents in biological terms the rudimentary vessel network which is formed before perfusion sets in. During this optimization we mark edges whose radius falls below a critical threshold $r^*_c$. These edges are no longer updated and are considered to have a radius of virtually zero (though for computational reasons they are here set to $r^*_c=10^{-21}$). We call such edges 'pruned' which corresponds to the biological phenomenon of having a vessel degenerate and collapse. Then we remove all pruned edges and disconnected vertices from the networks and calculate the remaining number of cycles according to equation \eqref{eq:cycles}. We then define the relative nullity of an equilibrated network,
\g{
\varrho=\frac{m-n+1}{z_0}
} 
as an order parameter, where $z_0$ is the initial number of independent loops before adaptation. Hence $\varrho=0$ corresponds to a treelike network while $\varrho>0$ captures the relative amount of redundancy in comparison to the initial plexus.
\section{ Numerical evaluation of the model framework }\label{sec:results}
In this section we present the simulation results acquired by solving the dynamical system \eqref{eq:radius_scaled_3} numerically until the system reaches a stationary state. Of particular interest is the final network's topology, i.e.\ its relative reticulation characterized by the order parameter $\varrho$.  We study in detail the dependence of $\varrho$ on the coupling $\lambda_1$, the volume penalty $\lambda_2$, and the fluctuation rate  $\lambda_3$. Primary focus lies on the interplay of coupling and fluctuation and how the underlying three dimensional lattice topology affects the remodeling process. All diagrams shown represent the results for one of the two intertwined networks; the results are symmetric for the other network due to a symmetric choice of the effective parameters $\lambda^i_1$, $\lambda^i_2$ and $\lambda^i_3$, see Appendix \ref{apx:simulations} for further details.\\
As underlying graph topology for the initial state networks we take the graph skeletons of the triply-periodic minimal surfaces P (`dual' simple cubic, see Figure~\ref{fig:model_b}), D (`dual' diamond cubic) and G (`dual' Laves) \cite{1996PhRvE..54.5012G}. These systems present highly symmetric and complementary space filling graphs and enable us to construct well defined intertwined networks with clear local edge affiliations. The initial edge radii are chosen randomly. We set the fluctuation rates identical $\lambda_3^1=\lambda_3^2$ as well as the coupling strength $\lambda_1^1= \lambda_1^2$. We did not find any qualitative differences in our results among different intertwined topologies. We thus present here the results for the simplest starting topology, the intertwined cubic lattices.
\subsection{Fluctuation induced nullity transitions independent of volume penalty and topology}
\begin{figure}[h]
	\hspace{-1.5cm}
		\begin{minipage}{0.375\columnwidth}
\subfloat[]{\label{fig:prune_1} \includegraphics[scale=.9]{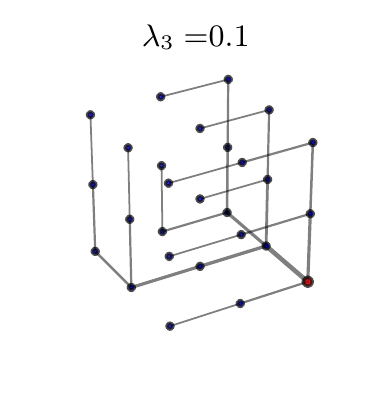}}\\
\subfloat[]{\label{fig:prune_3} \includegraphics[scale=.9]{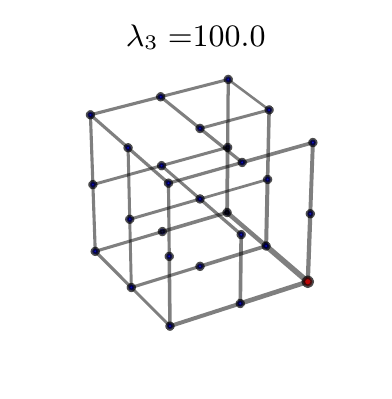}}
		\end{minipage}
			\begin{minipage}{0.6\columnwidth}
			 \subfloat[]{\includegraphics[scale=1.]{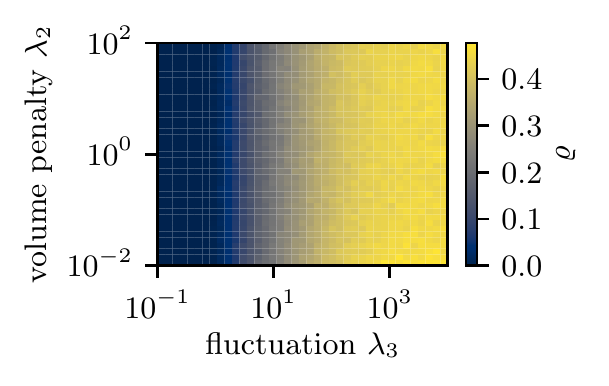}\label{fig:scale_a}}\\
	 \subfloat[]{\includegraphics[scale=1.]{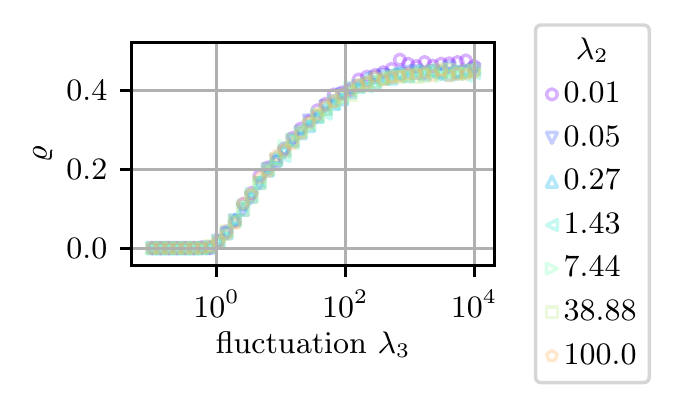}	\label{fig:scale_b}}
	 \end{minipage}
			\caption{Stationary network configurations for $\lambda_2=1$. The edge thickness is representative for the relative tube radius $\rho$. Sinks are marked as blue dots, the source as large red circle. \subR{fig:prune_1}~Tree-like for $\lambda_3=0.1$ \subR{fig:prune_3}~Fluctuation induced loops for $\lambda_3=100$.  \subR{fig:scale_a}+\subR{fig:scale_b}~Uncoupled adapting networks display continuous, logarithmic $\lambda_2$-independent nullity $\varrho$ transitions ($\lambda_1=0$, cubic lattice, $n=216$ and $m=540$). }\label{fig:scale}
		\end{figure}

First, we test the original Hu--Cai model in combination with Corson's fluctuation approach, i.e.\ the uncoupled case of $\varepsilon=0$ for three dimensional lattices. We do so for two reasons: First, to confirm the robustness of the adaptation mechanic for a plexus represented by non-planar graphs in a similar manner to \cite{Grawer:2015bh}. Second, to confirm the independence of the fluctuation-induced nullity transition from the volume penalty $\lambda_2$, as indicated in \cite{Hu:2013io}. To do so, we calculate the adaptation with a single corner source node (sinks otherwise) for a systematic scan of $\lambda_2$ and $\lambda_3$ (see Figure~\ref{fig:scale}).\\
Indeed, we recover the transition from tree-like configurations for $\lambda_3 \leq 1$ (Figure~\ref{fig:prune_1}) towards states exhibiting fluctuation induced loops for large $\lambda_3$  (Figure~\ref{fig:prune_3}). The emerging transition is of logarithmic nature, effectively saturating for $\lambda_3\approx 10^4$. In particular, we confirm that only an increase in the fluctuation ratio, $\lambda_3$, results in an increase in the nullity (Figure~\ref{fig:scale_a}). The continuous transition observed is independent of the system's effective volume penalty $\lambda_2$ (Figure~\ref{fig:scale_b}). Note however, that the parameter $\lambda_2$ influence the final vessel diameter as well as the time scales for reaching the stationary state (as does the system size if the number of identical sinks scales with system size). Further, we confirm that the adaptation mechanism reproduces the qualitative network topologies in three dimensional lattices as found before in planar graphs. From here we turn our attention to the fluctuation induced transition in comparison to the new spatial coupling.
	
\begin{figure*}[t]
  \subfloat[]{\label{fig:bilayer_nullity_a_rep} \includegraphics[scale=1.]{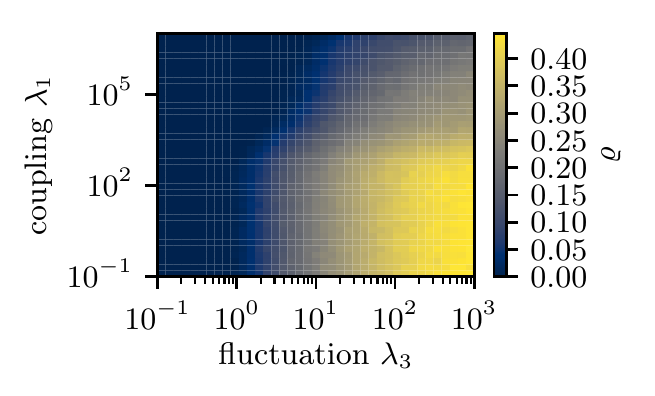}}\hspace{-0.5cm}
 \subfloat[]{	\label{fig:bilayer_nullity_b_rep} \includegraphics[scale=1.]{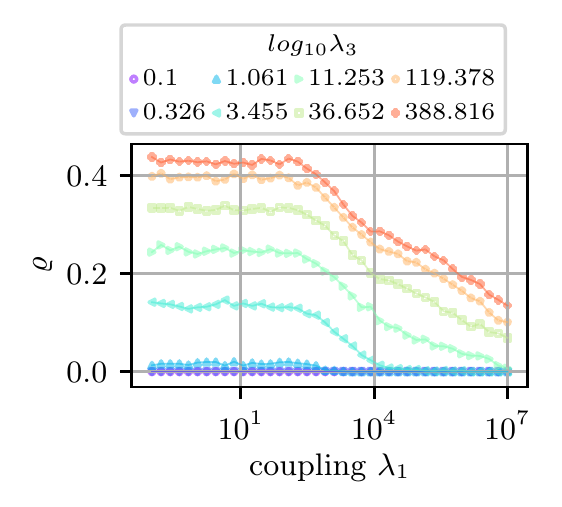}}\hspace{-0.5cm}
\subfloat[]{	\label{fig:bilayer_nullity_c_rep} \includegraphics[scale=1.]{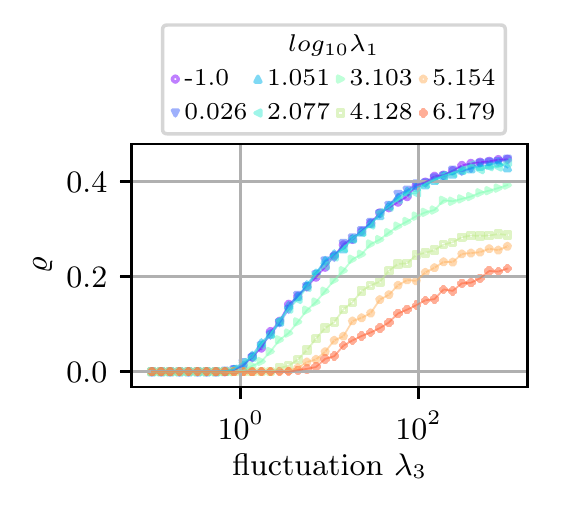}}\\

  \subfloat[]{	\label{fig:bilayer_nullity_d_rep} \includegraphics[scale=1.]{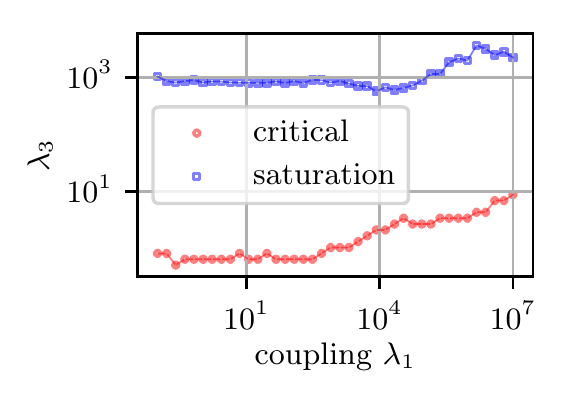}}\hspace{-0.5 cm}
  \subfloat[]{	\label{fig:bilayer_nullity_e_rep} \includegraphics[scale=1.]{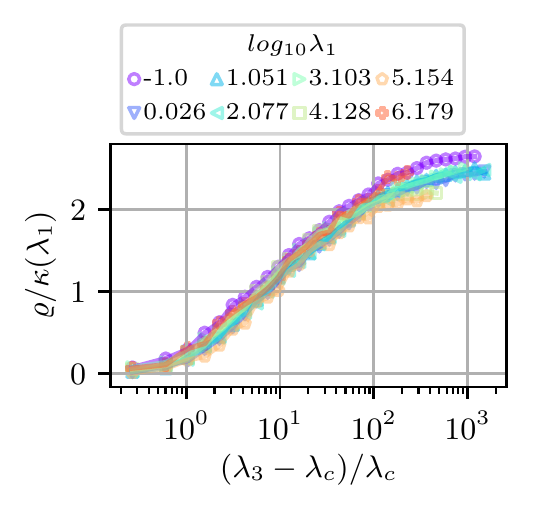}}\hspace{-0.5 cm}
  \subfloat[]{	\label{fig:bilayer_nullity_f_rep} \includegraphics[scale=1.]{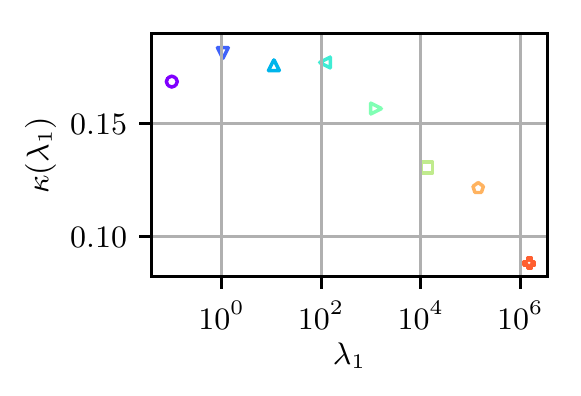}}
%
\caption{Nullity $\varrho$ transitions and state diagram for repulsively coupled networks ($\varepsilon=-1)$:  \subR{fig:bilayer_nullity_a_rep} Coupling vs fluctuation, $\bra{\lambda_1,\lambda_3}$-diagram with nullity $\varrho$ color map. \subR{fig:bilayer_nullity_b_rep} Coupling $\lambda_1$ induced nullity breakdown for varying fluctuation $\lambda_3$. \subR{fig:bilayer_nullity_c_rep} Fluctuation $\lambda_3$ induced nullity onset for varying coupling $\lambda_1$ \subR{fig:bilayer_nullity_d_rep} Critical $\lambda_3$ for onset of nullity transition for $\lambda_1$ variation and extrapolated saturation of nullity transition.
\subR{fig:bilayer_nullity_e_rep} Single curve collapse near fluctuation induced nullity onset $\lambda_c$ and scaled nullity $\varrho/\kappa(\lambda_3)$. We find the transition to be approximated by ${\varrho\bra{\lambda_1,\lambda_3}\approx \kappa \bra{\lambda_1} \bra{ \log_{10} \bra{\frac{\lambda_3-\lambda_c}{\lambda_c}} -1}}$
\subR{fig:bilayer_nullity_f_rep}$ \kappa \bra{\lambda_1} $ dependency derived from linear interpolation of  rescaled transition curves.
}\label{fig:bilayer_repulsive}
\end{figure*}	
\subsection{Repulsive coupling shifts and rescales fluctuation induced nullity transition}
\begin{figure}[h]
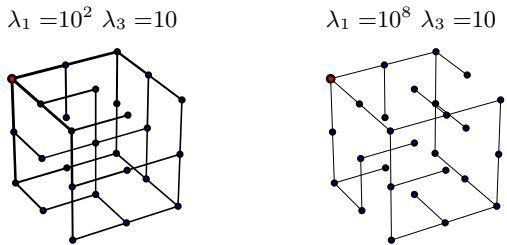

\input{./Figures_Main/28258889/N_32E_36_200000x5e-06mode_pointleaf_alpha_10.0_beta_100.0/eval/network_3d_plot_1.pgf}
 \input{./Figures_Main/28258889/N_32E_36_200000x1e-09mode_pointleaf_alpha_10.0_beta_100000000.0/eval/network_3d_plot_1.pgf}
    \caption{ Stationary network configurations for $\lambda_2=10^6$. The edge thickness is representative of the relative tube radius $r^*$. Sinks are marked as blue dots, the source as a large red circle. Repulsive coupling causes nullity breakdown and overall reduction in edge thickness leading to the disappearance of the usual vessel hierarchy.}	\label{fig:bilayer_skeleton_1} 
  	\end{figure}

To consider the novel, spatially coupled cases, we systematically scan the effective network coupling and flow-fluctuation parameters. In this section we will focus on the case of repulsive interactions, i.e.\ setting the coupling exponent to $\varepsilon=-1$. Of particular interest is the influence of the negative feedback this interaction introduces to the dynamical system \eqref{eq:radius_scaled_3}. \\
First, we see the fluctuation-induced nullity transition (as observed for the uncoupled system) to be preserved for weak couplings, $\lambda_1 < 10^4$. The full state diagram though (Figure~\ref{fig:bilayer_nullity_a_rep}) shows that the system's nullity may be influenced above that threshold not only by the rate of fluctuations $\lambda_3$, but also by the mutual repulsion of the two networks. Figure~\ref{fig:bilayer_nullity_b_rep} shows the corresponding $\lambda_1$ induced breakdown, displaying the possibility to nullify any reticulated structures by increasing the repulsive coupling strength in the system. Nevertheless, it seems that the influence of the repulsion is weaker in comparison, needing coupling parameters to be orders of magnitudes larger than the fluctuation rates. On the other hand, we also find the fluctuation-induced nullity onset to be continuous, as it was for the uncoupled system. Starting from a tree-like state at small fluctuations and increasing monotonically in a logarithmic manner beyond a critical $\lambda_3 \geq\lambda_c$ (Figure~\ref{fig:bilayer_nullity_c_rep}) we have the $\varrho$-trajectory eventually saturating for large fluctuation rates $\lambda_3> 10^3$ towards a maximal nullity $\varrho_{\text{max}}$. This leaves the network in a reticulated state, still displaying a visible vessel hierarchy towards the source. We can recover almost tree-like network states for increased repulsion rates $\lambda_1$, even losing the typical vessel hierarchy towards the source, see Figure~\ref{fig:bilayer_skeleton_1}. This increase in $\lambda_1$ further shifts the onset of nullity. To quantify these shifts we acquire the critical $\lambda_c$ by identifying the departure from zero in Figure~\ref{fig:bilayer_nullity_c_rep}. 
The critical point $\lambda_c$ seems to monotonically increase with the coupling parameter $\lambda_1$. Following up on this observation we extrapolated the onset of saturation in Figure~\ref{fig:bilayer_nullity_c_rep} by means of sigmoidal fits. The shifts of these indicators are shown in Figure~\ref{fig:bilayer_nullity_d_rep}, displaying a general increase of both the critical value and the saturation for increasing $\lambda_1$. Using the acquired critical values $\lambda_c$ we rescale the trajectories of Figure~\ref{fig:bilayer_nullity_c_rep} between the onset of the nullity transition and its saturation, as shown in Figure~\ref{fig:bilayer_nullity_e_rep}. Introducing the reduced fluctuation parameter $\frac{\lambda_3-\lambda_c}{\lambda_c}$ we find the trajectories to collapse on a single master curve, following a trivial logarithmic law as:
 \g{
 \varrho\bra{\lambda_1,\lambda_3}\approx \kappa\bra{\lambda_1} \bra{ \log_{10} \bra{\frac{\lambda_3-\lambda_c}{\lambda_c}} -1} \label{eq:log_law_rep}
 }
with the coupling dependent scale $\kappa$ acquired by interpolation of the data by equation \eqref{eq:log_law_rep}. We find $\kappa$ to be a decreasing function the coupling $\lambda_1$ as shown in Figure~\ref{fig:bilayer_nullity_f_rep}. This shows that the nullity breakdown and shift can be tuned for any given fluctuation rate by the coupling alone. Further, the negative feedback (caused by the repulsion of the two networks) does \textit{not} cause any shunting (i.e.\ collapse and disconnection of large sections of the networks) whatsoever. \\
For all simulations shown we set the response ${\lambda^1_0=\lambda_0^2=10^{-4}}$, and volume penalty ${\lambda^1_2=\lambda^2_2=10^6}$ (providing reasonable computation times for reaching stationary states and preventing the problem from becoming too stiff). The initial edge radii are chosen randomly and are subsequently continuously monitored to fulfill:
\g{
0<r^*_e+r^*_{e'}<1\label{eq:monitor}
} 
in order to prevent negative radii, or radii combinations corresponding to intersections.
\subsection{Attractive coupling induces a new nullity transition, fully recovering initial plexus}
\begin{figure}[h]
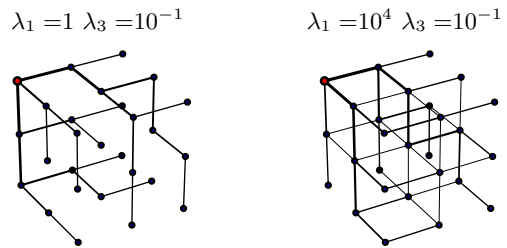

 \input{./Figures_Main/28258873/N_32E_36_300000x1e-06mode_pointleaf_alpha_0.1_beta_-1.0/eval/network_3d_plot_1.pgf}
   \input{./Figures_Main/28258873/N_32E_36_300000x1e-06mode_pointleaf_alpha_0.1_beta_-10000.0/eval/network_3d_plot_1.pgf}
   \caption{Stationary network configurations for $\lambda_2=10^6$. The edge thickness is representative for the relative tube radius $r^*$. Sinks are marked as blue dots, the source as a large red circle. Attractive coupling sets another local positive feedback which fixes overall nullity to the point of total plexus recovery. }  \label{fig:bilayer_skeleton_2}
		\end{figure}

\begin{figure*}[t]
 \hspace{-0.45cm}
  \subfloat[]{\label{fig:bilayer_nullity_a_att} \includegraphics[scale=1.]{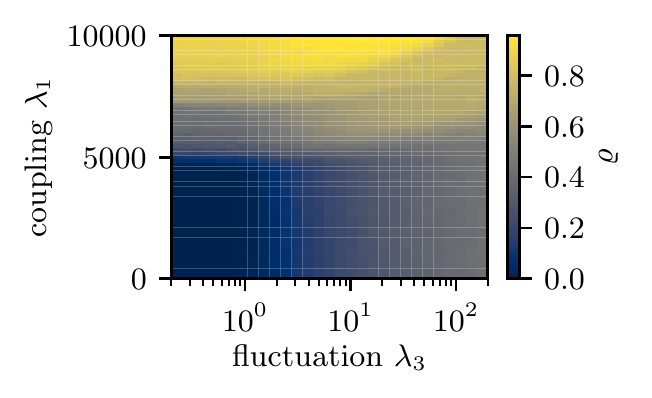}}\hspace{-0.45cm}
   \subfloat[]{	\label{fig:bilayer_nullity_b_att} \includegraphics[scale=1.]{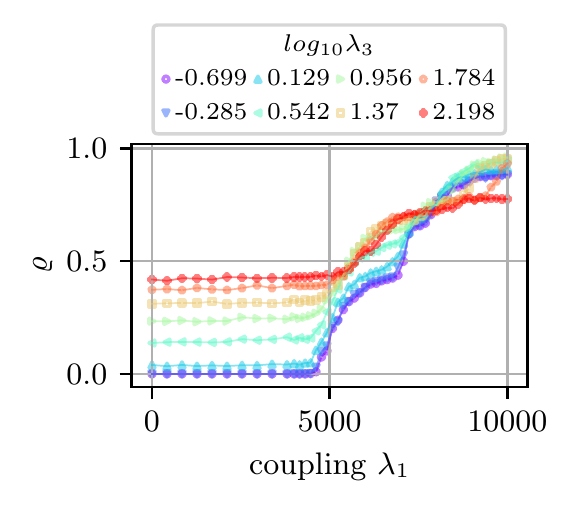}} \hspace{-0.45cm}
      \subfloat[]{	\label{fig:bilayer_nullity_c_att} \includegraphics[scale=1.]{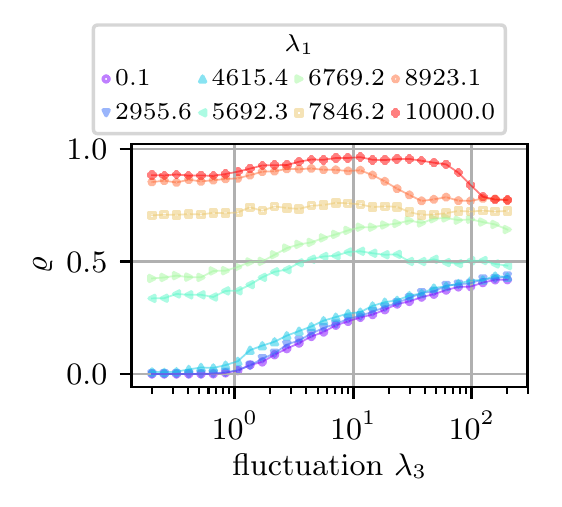}}\\
   \subfloat[]{\label{fig:bilayer_nullity_d_att} \includegraphics[scale=1.]{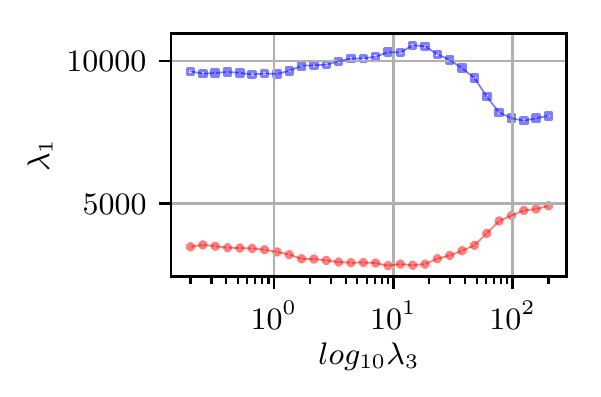}}\hspace{-0.5cm}
     \subfloat[]{	\label{fig:bilayer_nullity_e_att} \includegraphics[scale=1.]{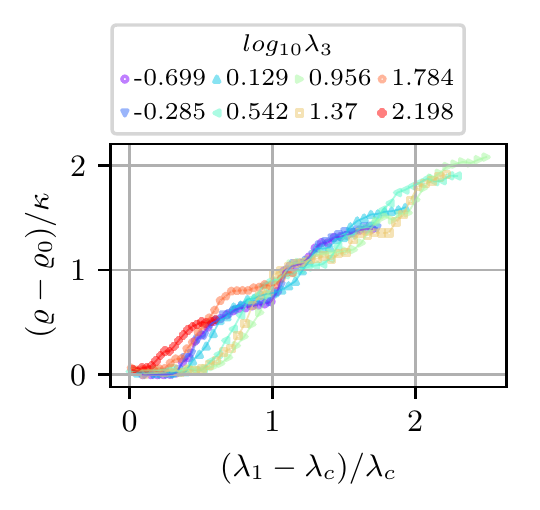}}\hspace{-0.5cm}
   \subfloat[]{\label{fig:bilayer_nullity_f_att} \includegraphics[scale=1.]{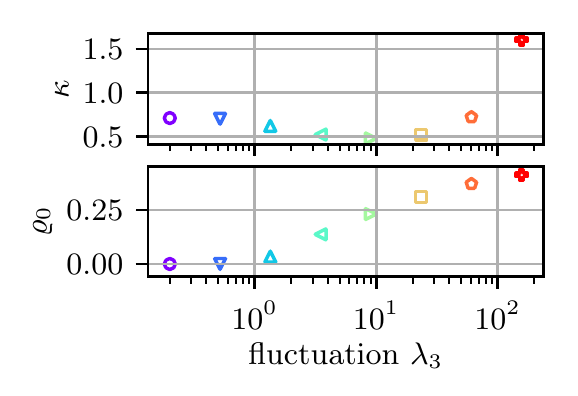}}
\caption{Nullity $\varrho$ transitions and state diagram for attractively coupled networks ($\varepsilon=3)$:   \subR{fig:bilayer_nullity_a_att} Coupling vs fluctuation, $\bra{\lambda_1,\lambda_3}$-diagram for nullity $\varrho$ color map.  \subR{fig:bilayer_nullity_b_att} Coupling $\lambda_1$ induced nullity transition for varying fluctuation $\lambda_3$. \subR{fig:bilayer_nullity_c_att}  Fluctuation $\lambda_3$ induced nullity transition for varying coupling $\lambda_1$.  \subR{fig:bilayer_nullity_d_att} Critical $\lambda_1$ for onset of nullity transition for $\lambda_3$ variation and extrapolated saturation of nullity transition.
\subR{fig:bilayer_nullity_e_att} Single curve collapse near coupling induced nullity onset $\lambda_c$ and scaled nullity $(\varrho-\varrho_0(\lambda_3))/\kappa(\lambda_3)$. We find the transition to be approximated by $\varrho\bra{\lambda_1,\lambda_3}\approx \kappa \bra{\lambda_3} \frac{\lambda_1-\lambda_c}{\lambda_c} +\varrho_0$
\subR{fig:bilayer_nullity_f_att} $\kappa$ and $\varrho_0$ dependency derived from linear interpolation of rescaled curves.
}\label{fig:bilayer_attractive}
		\end{figure*}

We now consider intertwined networks with an \textit{attractive} spatial coupling. We initialize the system with a positive coupling exponent, $\varepsilon =3$. We are particularly interested how the positive feedback this interaction introduces to the dynamical system \eqref{eq:radius_scaled_3} interacts and compares with the fluctuation induced nullity transition. \\
Indeed, the increased positive feedback from $\lambda_1$ leads to a significant increase in the system's redundancy, as can be seen in Figure~\ref{fig:bilayer_skeleton_2}. In Figure~\ref{fig:bilayer_attractive} we show the resulting nullity state diagrams and transition curves for a systematic scan the couplings and fluctuation rates. For increasing coupling strength $\lambda_1$ we see the emergence of a new nullity transition for $\lambda_1 > \lambda_c$. This transition recovers virtually the entire initial plexus for $\lambda_1 \rightarrow10^4$ (Figure~\ref{fig:bilayer_nullity_b_att}). Furthermore, this transition is significantly sharper than the fluctuation-induced case as it does not occur on a logarithmic scale. Any increase in fluctuations, $\lambda_3$, generates positive offset of the nullity curve (Figure~\ref{fig:bilayer_nullity_b_att}) indicating a constructive superposition of the mechanisms at hand. Yet, the trajectory's general form seems well preserved, while the saturation level is reduced for increased $\lambda_3$. Once again we determine the transition's onset and saturation regime, see Figure~\ref{fig:bilayer_nullity_d_att}. To do so for the onset, we calculate the trajectories' root of the onset after subtracting the trajectories' offset. The saturation regime is extrapolated via a sigmoidal fit. As in the previous section we are able collapse the trajectories onto a single master curve(Figure~\ref{fig:bilayer_nullity_e_att}) by following a trivial linear law :
\g{
\varrho\bra{\lambda_1,\lambda_3}\approx\kappa\bra{\lambda_3}\bra{\frac{\lambda_1-\lambda_c}{\lambda_c}} +\varrho_0\bra{\lambda_3} \label{eq:log_law_att}
}
with rescaled x-axis $\frac{\lambda_1-\lambda_c}{\lambda_c}$. The $\lambda_3$ dependency of the respective scaling parameters are shown in Figure~\ref{fig:bilayer_nullity_f_att}. We note here that these curves are slightly different from the complementary network in the case of cubic lattice topology as they are considerably more spread out, see the supplementary material \ref{apx:simulations} for details. This phenomenon does seem to be topology dependent, as is not present in the case of Lave-graphs or diamond lattices. The fluctuation induced nullity transition on the other hand is in some sense perturbed beyond the $\lambda_1$ transition, see Figure~\ref{fig:bilayer_nullity_c_att}. It seems an underlying competition of mechanisms is observable for large $\lambda_3$ as the level of saturation is reduced, as can be directly seen in the scale factor $\kappa$, see Figure~\ref{fig:bilayer_nullity_f_att}. 
Nevertheless, this interplay between different positive feedback mechanisms creates multiple nullity states, see Figure~\ref{fig:bilayer_nullity_a_att}, tuning the structures between spanning trees, partially reticulated and fully recovered plexus.  \\
For all simulations we set the response $\lambda_0^i=10^{-4}$, and volume penalty $\lambda_2^i=10^6$ as in the previous section. The parameter $\lambda_3$ is only considered here for $\lambda_3 <300$ as affiliated edge pairs will violate the contact condition \eqref{eq:monitor} beyond this range.

\section{Generalizing Murray's law }\label{sec:estimation}
Our model of spatial coupling also points to a new form of geometric law at vessel branchings. Recall Murray's Law, which connects the radii $r_0$ of a parent vessel splitting into at least two child branches with radii $r_1$, $r_2$ as:
\alg{r_0^{\alpha}=r_1^{\alpha}+r_2^{\alpha} \label{eq:murray1}}
In the original formulation an exponent of ${\alpha=3}$ was predicted as the outcome of a cost optimization process \cite{Murray:1926tj} which relates directly to the dissipation-volume minimization procedure discussed in section \ref{sec:cost_ansatz}. Further, rescaled cost models \cite{Bohn:2007fi} which consider cost variations via an exponent ${c_e \rightarrow c_e^{\gamma}}$ suggest 
\alg{\alpha=2\bra{\gamma+1} \label{eq:gamma}}
while discarding flow fluctuations. We illustrate here the problems one encounters when testing these power laws for real intertwined structures such as sinusoids and bile canaliculi in the mammalian liver. Further, we introduce a new generalized form of Murray's Law which takes into account fluctuations as well as geometric coupling.  We then use this new form to estimate the interaction parameters of the real system heretofore inaccessible to experimental investigation.\\
The datasets were acquired from collaborators at the MPI-CBG in the following way: Mouse livers from adult mice were fixed by trans-cardial perfusion, sectioned into 100 mm serial slices, optically cleared and immunostained,  as described in \cite{MoralesNavarrete:2015kk}. To visualize the different tissue components, the tissue sections were stained for nuclei (DAPI), cell borders (Phalloidin), bile canaliculi network (CD13), and the extracellular matrix (ECM, fibronectin and laminin) facing the sinusoidal network \cite{MoralesNavarrete:2019ku}. High-resolution images of the liver lobule (Central vein – portal vein axis) were acquired by using confocal microscopy with a 63x/1.3 objective ($0.3\ \mu m$ voxel size). Finally, the resulting images were segmented and network skeletons calculated with the Motion Tracking software as described in \cite{MoralesNavarrete:2015kk} and \cite{MoralesNavarrete:2016gb}.
\subsection{ Classical Murray's Law inadequate for reticulated, intertwined vessel structures of the liver lobule }
\begin{figure}[h]
\includegraphics[scale=1.]{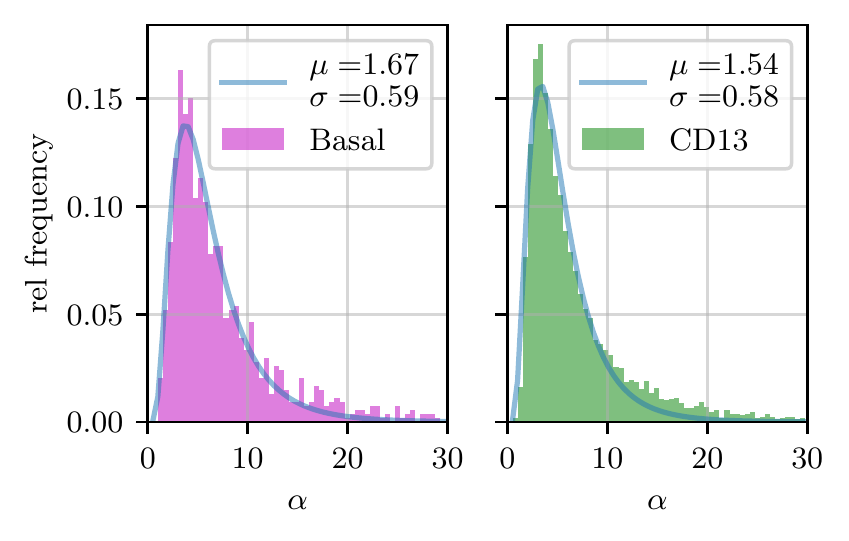}
\caption{
Exponent distributions found for Murray's Law \eqref{eq:murray1} and respective log-normal fits: Sinusoidal systems (basal marker) with mode $\alpha=e^{\mu-\sigma^2}\approx 3.75$ and bile canaliculi systems (cd13 marker) with mode $\alpha=e^{\mu-\sigma^2}\approx 3.33$.
 }  \label{fig:murray}
\end{figure}

We test Murray's Law for the network skeletons of the sinusoids and bile canaliculi in the liver lobule by fitting the exponent $\alpha$ in equation \eqref{eq:murray1} for every branching of degree three (Y-branching).  There is significant deviation from the predicted exponent $\alpha=3$  (Figure~\ref{fig:murray}). 
The modes of the acquired, log-normally distributed fit exponents are $\alpha\approx 3.75$ for sinusoids and for bile canaliculi $\alpha\approx 3.33$. As capillary systems were already known to defy the cubic relationship \cite{Sherman:1981ux}, we suspect this mismatch to be correlated with the reticulated nature of these network types. Further, one expects other mechanisms than mere dissipation-volume minimization will be involved, making these deviations not well described by the cost exponent $\gamma$ alone. Indeed, in accordance to equation \eqref{eq:gamma} one would deduce $\gamma < 1$ for the given liver lobule datasets which is in direct contradiction with the rescaled cost model \cite{Bohn:2007fi}, which predicts an exponent $\gamma$-induced nullity transition only for $\gamma\geq1$. On the other hand this could potentially be circumvented if fluctuation-induced loops are considered as well \cite{Katifori:2010kt,Corson:2010ee}. However, to our knowledge it has not yet been discussed how such fluctuations alter Murray' Law.\\ We deduce from our pruning model a new set of coefficients, $a_e$, which are dependent on their corresponding edge's neighborhood and the respective coupling strength, as well as the global structure of sinks and sources (which were assumed to be uncorrelated and identically distributed). This procedure greatly alters the form of equation \eqref{eq:murray1} and we derive a new geometric law derived from the steady-states of the ODEs in \eqref{eq:radius_scaled_3}. We recover the cubic exponent of the original model with new coefficient corrections that depend on the strength of fluctuations and spatial coupling:
 \alg{\sum_e a_e r_e^{*,3} & =1 \label{eq:murray2}\\
 \text{ with } a_e & =\pm\sqrt{\frac{\lambda_2 - \lambda_1 r_e^{*,-1}g_e^*}{1+\lambda_3\frac{\delta\phi_e^*}{\phi_e^*}}}\nonumber}
The effective coupling $g_e^*=\sgn(\varepsilon)\sum_{e'}F_{ee'}\LP r_{ee'}^{*\varepsilon-1}$, and squared edge potentials $\phi^*_e$, $\delta\phi^*_e$ are defined as in section \ref{sec:theory}. This new law \eqref{eq:murray2} may be further generalized in case of more complicated flow landscapes, see Appendix \ref{apx:scaling laws}. 
\subsection{High accuracy prediction of interaction parameters for  ideal Kirchhoff networks}\label{subsec:ideal}
\begin{figure*}[t]
\subfloat[Mono Source]{\label{fig:murray_par1} \includegraphics[scale=1.]{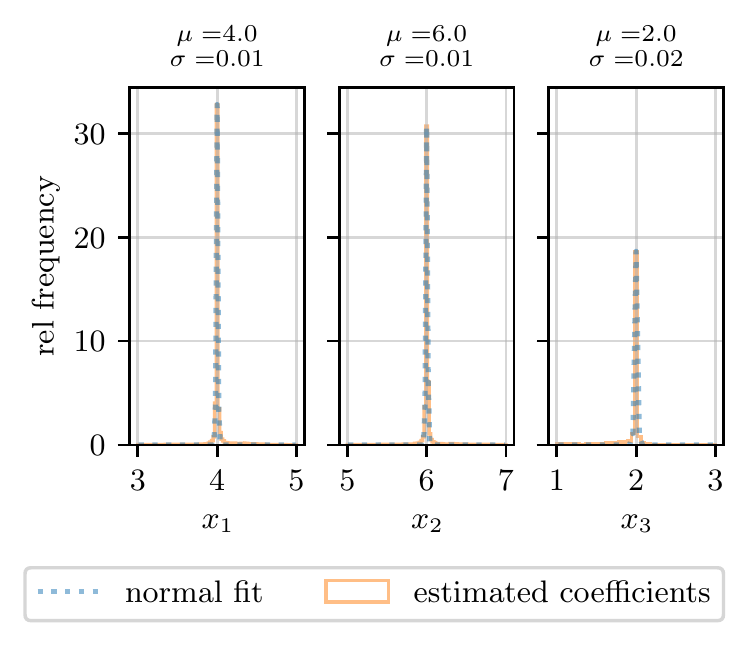}}
\subfloat[Multiple Sources]{\label{fig:murray_par5} \includegraphics[scale=1.]{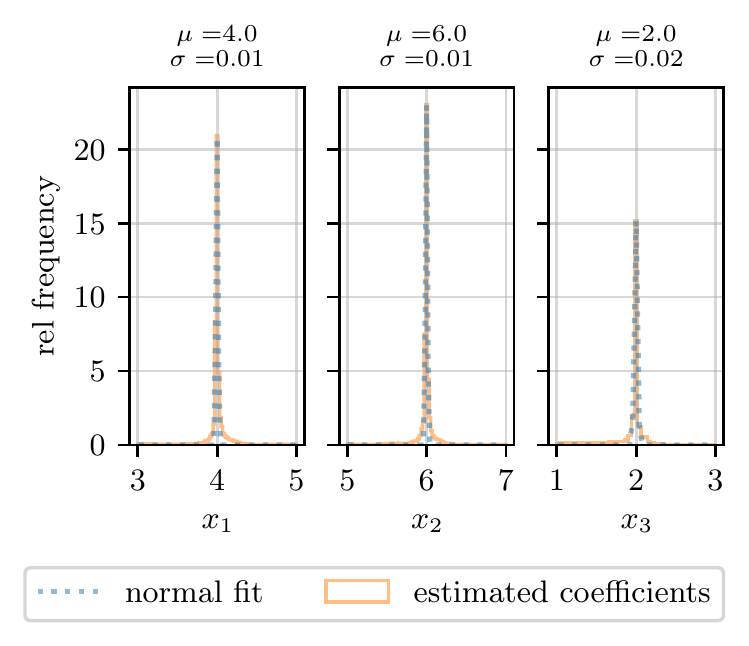}}
\caption{Test of coupled ideal Kirchhoff networks, dual Laves-Graph topology, rescaled axis is $x_i=\log_{10}\lambda_i$, the system was initialized symmetrically with ${x_1 =4}$ , ${x_2 =6}$ and ${x_3 =2}$ for the repulsive case $\varepsilon =-1$. 
\subR{fig:murray_par1}: Single source/multi-sink system, parameters histograms and fits $\mathcal{N} (\mu,\sigma)$ presenting the estimated simulation parameters $\lambda_i$. \subR{fig:murray_par5}: Multi-source/multi-sink system, parameters histograms and fits $\mathcal{N} (\mu,\sigma)$ presenting the estimated simulation parameters $\lambda_i$. 
 }
 \label{fig:par1}
 \end{figure*}

We tested the feasibility of the new generalized Murray's Law \eqref{eq:murray2} by simulating the pruning on a dual Laves graph topology (3-regular), with $n=206$ ($n'=210$) vertices and $m=259$ ($m'=261$) edges and setting the parameters symmetrically to $\lambda_1^i=10^4$, $\lambda_2^i=10^6$, $\lambda_3^i=10^2$. The sources were positioned in random vertices of the system. Edges of the respective networks were affiliated with each other by finding the nearest neighbors of edges inside a perimeter $\delta$. We numerically \cite{SciPy} find the roots of equation \eqref{eq:murray2} for a set of positive definite $\lambda_1^i$, $\lambda_2^i$, $\lambda_3^i$. As we do not intend to use information on the direction of the currents at the sink-nodes (as this information is not available in the real system) we solve equation \eqref{eq:murray2} for the seven relevant sign permutations at each branching, see Appendix \ref{apx:scaling laws}, Figure~\ref{fig:floworder}. For further evaluation only the fit of highest quality (function value) is used. We use a logarithmic rescaling $x_i=\log_{10}\lambda_i$ in order to find a symmetric representation of the histogram's data. Doing so we fit a normal distribution $\mathcal{N}(\mu,\sigma)$ to the histogram's maxima and we find strong agreement with the actual parameters for both networks, see Figure~\ref{fig:murray_par1}  (depicted here only for the case of repulsive coupling $\varepsilon =-1$). We also tested for an abundance of multiple sources distributed throughout the system (acting as identical copies of each other). We find our procedure correctly recovers the relevant model parameters in these cases as well (Figure~\ref{fig:murray_par5}). We found in the same way good agreement for the case of attractive coupling.
\subsection{Limited estimation of interaction parameters for sinusoidal networks}\label{subsec:estimating_real}
\begin{figure}[t]
\subfloat[Attractive Coupling]{\includegraphics[scale=1.]{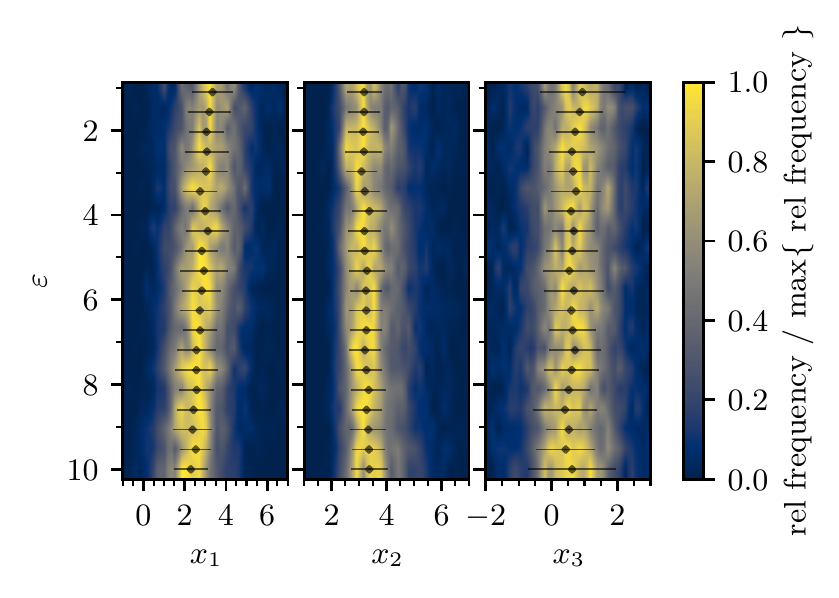}\label{fig:murray_par2} }\\
\subfloat[Repulsive Coupling]{\includegraphics[scale=1.]{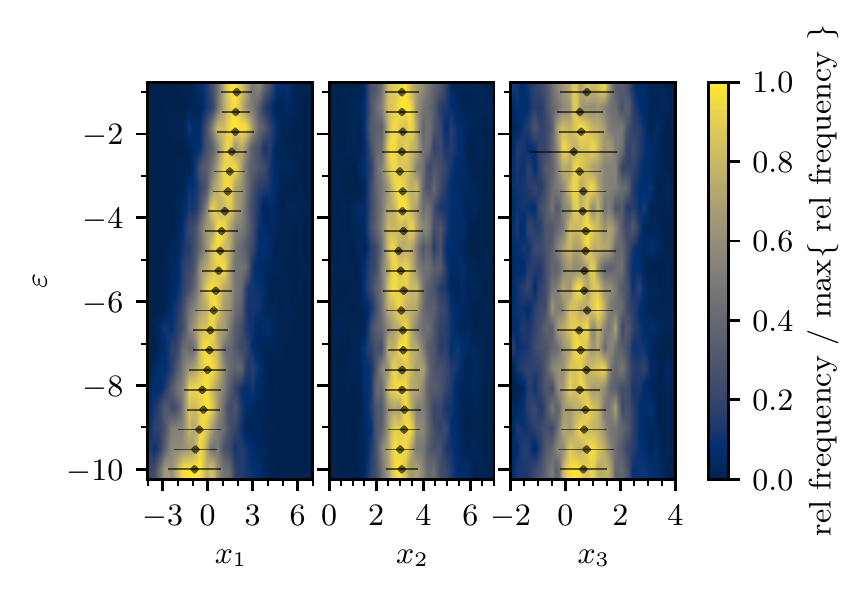}\label{fig:murray_par3} }
\caption{Estimated parameters for the sinusoidal system near the central vein for $R=397\ \mu m$, $\delta= 30\ \mu m$ and rescaled axis $x_i=\log_{10}\lambda_i$ for different coupling exponents $\varepsilon$. }\label{fig:murray_par_fit}
\end{figure}

Finally we use the extracted graph skeletons of the sinusoids and bile canaliculi in the vicinity of a central vein in the mouse liver to test the generalized Murray's Law on a real intertwined vessel system. The vascular structure is represented by sets of vertices and edges bearing the positional and radial information of the respective vasculature. We use the same approach as in the prior section to estimate the coupling $\lambda_1$, volume penalty $\lambda_2$ and fluctuation rate $\lambda_3$ for the sinusoidal system. We stress that no information about the actual flow nor the exact point of flow injection or drainage was available. Hence, we first make some simplifying assumptions about the positions of sinks and sources as well as cropping the network skeletons to reduce their complexity: First we identify the vertices in the sinusoidal network which are closest to the central vein and label them as sinks. The geometric center of mass of these vertices is calculated and used as the center of a sphere of radius $R$, representing the range of interest. Any other components, vertices or edges of any network positioned outside this perimeter are discarded (here $R=390\ \mu m$, in order to keep the resulting graphs at a moderate size for computational purposes). Next, all branching points in the sinusoidal network are identified, as are all paths $\vc{p}=\bra{e_i,...,e_j}$ consisting of edges $e_i$ which start from these points. We proceed for the canaliculi the same way and check for each segment of a path whether there is another segment of another network's path inside a perimeter $\delta$ (here chosen as $30\ \mu m$). If so, these paths count as affiliated. We merge all edges along a path into a single edge by using the conventional addition theorems for series of conductivities in Kirchhoff networks. Proceeding like this we end up with a reduced sinusoidal network, with $n=318$ vertices and $m=452$ edges. For further details on the reduction procedure see Figure~\ref{fig:sketch} in Appendix \ref{apx:scaling laws}.\\
We find the parameters $\lambda_1$, $\lambda_2$, $\lambda_3$ numerically by solving equation \eqref{eq:murray2} for randomly sampled tuples of branching points. We do so for a range of exponents $\varepsilon$, under consideration of all sign permutations for $a_j$. Hence we probe the system in different coupling regimes and consider all possible flux patterns. The solutions' histograms as well as the calculated means $\mu\bra{x_i}$ and standard deviations $\sigma\bra{x_i}$ are presented in Figure~\ref{fig:murray_par_fit}, again using the logarithmic axis rescaling $x_i=\log_{10}\lambda_i$. We find the histograms to be considerably broadened single peak distributions, indicating relatively stable means for $x_2$, $x_3$ with varying means of $x_1$ for variable exponents $\varepsilon$.\\
From the previous section \ref{sec:results} we concluded that attractively coupled networks ($\varepsilon >1$) are able to generate robust, reticulated structures for increased coupling rates of $\lambda_1$ and are supported in this by increased fluctuation rates $\lambda_3$. The estimates for the attractive coupling case in Figure~\ref{fig:murray_par2} indicate that increased fluctuation rates are present ($\mu\bra{x_3} >0$) which may account for reticulated structures in flow driven adaption. Further we observe monotonically increasing rates of coupling for increasing $\varepsilon$. Interestingly we find the coupling rates poised just below the actual onset of the topological transition e.g.\ $\mu\bra{x_1} \pm \sigma\bra{x_1}= 2.56 \pm 1.04$ for the coupling exponent $\varepsilon=3$.\\
Further we find the repulsive coupling case to reproduce the same regime of values for $x_2$, $x_3$, indicating reticulation by flow fluctuation. Yet the coupling parameter $\mu\bra{x_1}$ displays a monotonically decreasing behavior for decreasing values of $\varepsilon$, see Figure~\ref{fig:murray_par3}. Those low values of $x_1$ suggest repulsive interactions to be negligible as the estimates lie far from the regimes with topological implication, e.g.\ $\mu\bra{x_1} \pm \sigma\bra{x_1}= 1.94 \pm 1.03$ for the coupling exponent $\varepsilon=-1$.\\
Unfortunately all estimates $\mu\bra{x_i}$ are accompanied by large standard deviations, which in the logarithmic context, span orders of magnitudes. Further we find no indication for a specific coupling scenario $\varepsilon$, e.g.\ based on a collapse of the standard deviation for a specific $\varepsilon$, parameter distributions contradicting topological structure, etc. We suspect these issues to originate from several sources: segmentation inaccuracies' during image analysis, crude approximation of the sink-source landscape of the system, the chosen algorithm of complexity reduction and ambiguity of numeric solutions due to the non-linearity of the problem. Ultimately the very fact that we only make an educated guess about the adaptation mechanisms, might exclude other essential principles of self-organized vessel adaptation in the liver lobule. Nevertheless, considering these findings and the restrictions of our model's approach we assume the emergence of reticulated sinusoidal structures rather to be the product of flux fluctuations rather than of the newly proposed geometrical interactions. With this technique we have shown that it is possible to extract order of magnitude \textit{estimates} of otherwise inaccessible parameters of real adapting biological networks, and in doing so to be capable of making qualitative statements about the \textit{relative} strength or importance of different feedbacks.

%
\section{Discussion}\label{sec:discussion}
We have shown that spatial coupling presents another potential mechanism of controlling the topological complexity of optimal transport systems in 3D. We considered the special case of 'single source/multiple sinks' in combination with simple cubic lattices as plexi as the simplest case possible. In the case of repelling networks we find this interaction to reduce the networks' relative loop density and provide another method by which a system may be tuned towards its final architecture. It's also possible to retrieve tree-like states at increased fluctuation levels, which represents a new stabilization mechanism for spanning trees in noisy networks. Nevertheless, the onset of redundancy is primarily driven by the existence of flow fluctuations. On the other hand, we have shown that attracting vessel surfaces allow for the robust emergence of loops on their own. Even the full recovery of the initial plexus is feasible beyond the coupling induced nullity transition, in contrast to what is possible with fluctuations. This presents a new mechanism which allows for the maintenance of dissipation minimized reticulated vessel networks as long as there is an effective scaffold providing a positive growth feedback. No qualitative differences could be found in the phase diagrams in comparison with the dual diamond, or dual Laves graphs. Our model may be interpreted as a toy-model for intertwined flow networks as found in the mammalian liver lobules and other organ structures, such as kidneys, pancreas or bone marrow. The cost function ansatz, though, provides a generally applicable tool in network optimization, and could profitably applied to other boundary conditions or graph geometries which resemble realistic structures. \\ 
Our approach further enabled us to derive a more general form Murray's law, directly involving flow interactions and environmental couplings. We find this technique to predict the model parameters with high fidelity for simulated Kirchhoff networks given their topology and respective edge radii distributions. In the same manner we find order of magnitude estimates for the parameters in experimentally acquired data sets of sinusoids in liver lobule of mice. Hence one could consider this methodology as an effective classification of spatially adapting network structures allowing to probe for relevant parameter regimes of adaptation models.\\
We aim to expand the purely geometric approach which was studied here, by explicit involvement of hydrodynamic-chemical feedback between a vessel and its local environment. This would be of particular concern in intertwined distribution networks transferring water and metabolites with their respective partner network. Biological systems such as the capillary networks in the liver lobule, present complex dual transport systems where the actual flow rates are not necessarily influenced by their respective partner network's flow rate \cite{BRAUER:1954cj} but rather by the concentration of bile salt components transported \cite{Boyer:2013hj} as well as secretion rates of hepatocytes. Subsequently, future studies should consider concentration/pump-rate dependent flow rate perturbations in the optimization model. One may also consider a direct postulation of cross-network feedback in the adaptation dynamics to account for the actual network structures. Recent studies regarding solute transport and optimization of cross-wall transport \cite{Meigel:2018hw,Ostrenko:2019iu} might present a suitable basis for such an approach. Eventually, we strive towards a generalized formulation of environmental factors whose influence can be depicted in the form of local adaptation rules of complex flow networks.  Ultimately, complex distribution and flow systems that respect and leverage their spatial embeddings remain a deeply rich topic with myriad opportunities both to make contact with applied and biological settings and to open up new ways to understand the physics of complex systems.  

\begin{acknowledgements}
Our thanks go to Hernan Morales-Navarrete (image reconstruction) and Fabian Segovia-Miranda (experiments) for providing us with the sinusoidal and canaliculial skeleton datasets. We'd like to thank Marino Zerial, Julien Delpierre, Quentin Vagne , Dora Tang, Andre Nadler, Mark Warner and all members of the Modes Lab and Zechner Lab for their comments, helpful discussions and resourceful feedback  throughout the process of creating this work.
\end{acknowledgements}
\bibliography{../../../lib2} 
\newpage
\appendix
\section{Minimization of custom Lyapunov function $\Gamma$}\label{apx:gradient_flow}
To minimize the cost function $\Gamma$ as given in section \ref{sec:theory} we write it as
\alg{\Gamma & = \Gamma_{12}+\Gamma_1 + \Gamma_2\\ \text{ with } \Gamma_1 & =\frac{\pi}{8\eta_1 l_1}\sum_{e_1} \brc{\LP p_{e_1}^2 r_{e_1}^4+a_1 r_{e_1}^{2}} \\ \text{ and } \Gamma_2 & =\frac{\pi}{8\eta_2 l_2}\sum_{e_2} \brc{\LP p_{e_2}^2r_{e_2}^4+a_2 r_{e_2}^{2}}}
 We may further use the vectorial notation for dissipation-volume terms $\Gamma_1$, $\Gamma_2$, using \eqref{eq:trafo}, to formulate it in terms of the nodal sinks/sources (here just for $\Gamma_1$, derivation for $\Gamma_2$ is performed analogously),
 \alg{  \Gamma_1 &=\LP \vc{p}^T\cdot\vc{C}\cdot \LP\vc{p}+q\text{ }Tr\bra{\vc{C}^{1/2}}\\ &=\vc{s}^T\brc{\vc{B}^{T}\vc{C}\vc{B}}^{\dagger}\vc{s}+q\text{ }Tr\bra{\vc{C}^{1/2}}}
 with $q=a_1\bra{\frac{\pi}{8\eta_1 l_1}}^{1/2}$ and $c_{e_1}=\frac{\pi r_{e_1}^4}{8\eta_1l_1}$ as entries of the diagonal $\vc{C}$. We calculate the (pseudo-)time derivatives of P to be
		\alg{\frac{d\Gamma_1}{dt} & =\vc{s}^T\partd{t}\brc{\vc{B}\cdot \vc{C} \cdot \vc{B}^{T}}^{\dagger}\vc{s}\nonumber\\& +2\vc{s}^T\brc{\vc{B}\cdot \vc{C}\cdot \vc{B}^{T}}^{\dagger}\partd{t}\vc{s}+q\partd{t} Tr\bra{\vc{C}^{1/2}}}
		The derivative of the generalized inverse  ${\vc{B}\cdot \vc{C}\cdot\vc{B}^T =\vc{A}}$ being \cite{1973SJNA...10..413G},
		\alg{\partd{t} \vc{A}^{\dagger} & =-\vc{A}^{\dagger}\bra{\vc{B}\cdot \partd{t}\vc{C}\cdot\vc{B}^T} \vc{A}^{\dagger}\nonumber\\
		& +\brc{\vc{I}-\vc{A}^{\dagger}\cdot\vc{A}}\bra{\vc{B}\cdot \partd{t}\vc{C}\cdot\vc{B}^T}\vc{A}^{T\dagger}\cdot\vc{A}^{\dagger}\nonumber\\
		& +\vc{A}^{\dagger}\cdot\vc{A}^{T\dagger}\bra{\vc{B}^T\cdot \partd{t}\vc{C}\cdot\vc{B}}\brc{\vc{I}-\vc{A}\cdot\vc{A}^{\dagger}}}
		Fortunately the projector terms vanish as we have,
		\g{\vc{s}^T\brc{\vc{I}-\vc{A}^{\dagger}\cdot\vc{A}}=0 \text{ and } \brc{\vc{I}-\vc{A}\cdot\vc{A}^{\dagger}}\vc{s}=0}
		Together with the identity $\vc{A}^{\dagger}=\bra{\vc{A}^T}^{\dagger}$ the total time-derivative of $\Gamma_1$  becomes,
		\alg{\frac{d\Gamma_1}{dt} & =-\vc{s}^T\vc{A}^{T\dagger}\bra{\vc{B}\cdot\partd{t}\vc{C}\cdot\vc{B}^T}\vc{A}^{\dagger}\vc{s}\nonumber\\& +2\vc{s}^T\vc{A}^{\dagger}\partd{t}\vc{s}+q\partd{t}Tr\bra{\vc{C}^{1/2}}}
	With partial derivatives simplifying this formula as:
		 \alg{  \partd{t} Tr\bra{\vc{C}^{1/2}} & =\frac{1}{2} Tr\bra{ \vc{C}^{-1/2}\cdot\partd{t}\vc{C}}\\ \partd{t}\vc{s} & =0}
As we also have $\LP\vc{p}=\vc{B}^T\cdot\vc{A}^{\dagger}\vc{s}$, we may write the total time-derivative as 		
		\alg{\frac{d\Gamma_1}{dt}= -\LP\vc{p}^T\cdot\partd{t}\vc{C}\cdot\LP\vc{p}+\frac{q}{2} Tr\bra{\vc{C}^{ -1/2}\partd{t}\vc{C}}}
	 With diagonals $\partd{t}c_{e_1}=\frac{\pi r^3_{e_1}}{2\eta_1l_1}\partd{t}r_{e_1}$ and re-substituting $q$ this becomes
		\alg{\frac{d\Gamma_1}{dt}=\frac{\pi}{2\eta_1l_1}\sum_{e_1} r_{e_1}\brc{ \frac{a_1}{2} - \LP p_{e_1}^2r_{e_1}^2}\partd{t}r_{e_1}\label{eq:p_derivative}}	 
	 The coupling component's derivative $\Gamma_{12}$ may be calculated with partial derivatives, 
	 \alg{\frac{d \Gamma_{12}}{dt}=-\frac{b \varepsilon}{2}\brc{\sum_{e_1e_2}F_{e_1e_2}\LP r_{e_1e_2}^{\varepsilon-1}\partd{t}r_{e_1}+\sum_{e_1e_2}F_{e_1e_2}\LP r_{e_1e_2}^{\varepsilon-1}\partd{t}r_{e_2}}\label{eq:e_derivative}}
		so combining \eqref{eq:p_derivative} and \eqref{eq:e_derivative} we get for the total derivative of $\Gamma$,
		\alg{
		\frac{d\Gamma}{dt} & = \frac{\pi }{2\eta_1 l_1}\sum_e r_{e_1}\brc{ \frac{a_1}{2}-\LP p_{e_1}^2r_{e_1}^2}r_{e_1}\partd{t}r_{e_1}\nonumber\\&+\frac{\pi }{2\eta_2 l_2}\sum_{e_2} \brc{\frac{a_2}{2}-\LP p_{e_2}^2r_{e_2}^2}r_{e_2}\partd{t}r_{e_2}\nonumber\\ & -\frac{b\varepsilon}{2}\brc{\sum_{e_1}F_{e_1e_2}\LP r_{e_1e_2}^{\varepsilon-1}\partd{t}r_{e_1}+\sum_{e_2}F_{e_1e_2}\LP r_{e_1e_2}^{\varepsilon-1}\partd{t}r_{e_2}}
		}
	  To find a local minimum of $\Gamma$ when progressing through (pseudo-)time $t$ we have to ensure that $d\Gamma\leq 0$. So we may minimize $\Gamma$, with auxiliary functions $\chi_i\geq 0$ by choosing
	  \alg{
	  \partd{t}r_{e_1}/\chi_1& =\brc{\LP p^2_{e_1}r_{e_1}^2-\alpha_1}r_{e_1}+\beta_1 \sgn(\varepsilon)\sum_{e_2}F_{e_1e_2}\LP r_{e_1e_2}^{\varepsilon-1} \\ \partd{t}r_{e_2}/\chi_2 & =\brc{\LP p^2_{e_2}r_{e_2}^2-\alpha_2}r_{e_2}+\beta_2 \sgn(\varepsilon)\sum_{e_1}F_{e_1e_2}\LP r_{e_1e_2}^{\varepsilon-1}
	  } 
where we set ${\alpha_i=\frac{a_i}{2}}$, ${\beta_i=\frac{b\eta_i l_i\vert\varepsilon\vert}{\pi }}$ for compact notation.
	 
\section{Uncorrelated and coupled flow fluctuations}\label{apx:flow_fluctuation}
 In this section we give a detailed derivation of the analytic form of the mean squared pressure in case of uncorrelated, identically distributed sink fluctuations as introduced in \cite{Corson:2010ee} and discussed in section \ref{sec:theory}. \\ From the current law \eqref{eq:current_law} in combination with Ohm's law \eqref{eq:ohm_law} one knows that the sum over all in and outflows of the system vanishes \cite{DUFFIN:1959ti}, i.e.\ $\sum_v s_v=0 $. When considering the sink conditions as well as the source constraint defined in the theory section, we may write the moments as (putting the distinguished source at $v=0$),
		\alg{\mean{s_v} & =\mu \\ \mean{s_0} & =-\mean{\sum_{v>0}s_v}=\bra{1-n}\mu\label{eq:source1}\\ \mean{s_v s_w} & =\delta_{vw}\sigma^2+\mu^2 \text{ if }v,w>0\\  \mean{s_0^2} & =\mean{\sum_{v,w>0}s_vs_w}= \bra{n-1}\sigma^2 + \bra{n-1}^2\mu^2 \label{eq:source2}\\ \mean{s_0s_w} & =-\mean{\sum_{v>0}s_vs_w} =-\bra{n-1}\mu^2-\sigma^2 \label{eq:source3}}
Hence we may calculate the squared-mean pressure by using the auxiliary conductivity tensor ${\vc{A}^e_{vw}=\vc{C}^{-1}_{ee}\brc{\vc{B}\cdot\vc{C}^{1/2}}_{ev}^{\dagger}\brc{\vc{B}\cdot\vc{C}^{1/2}}_{ew}^{\dagger} } $ as,
		\alg{\mean{\LP p_e^2}   = &\sum_{vw}\vc{A}^e_{vw}\mean{s_v s_w}\label{eq:sq_p}\\
		 = &\vc{A}^e_{00}\mean{s_0^2}+2\sum_{v>0}\vc{A}^e_{v0}\mean{ s_0s_v } + \sum_{v,w > 0}\vc{A}^e_{vw}\mean{s_vs_w}\\
		   = &\vc{A}^e_{00}\bra{n-1}\brc{\bra{n-1}\mu^2+\sigma^2}\nonumber\\&-2\sum_{v>0}\vc{A}^e_{v0}\brc{\bra{n-1}\mu^2+\sigma^2} \nonumber\\& + \sum_{v,w > 0}\vc{A}^e_{vw}\brc{\delta_{vw}\sigma^2+\mu^2}\\
		    = & \sum_{v,w}\vc{A}^e_{vw}\bra{n-1}\brc{\bra{n-1}\mu^2+\sigma^2}\delta_{0v}\delta_{0w}\nonumber \\  &-\sum_{v,w}\vc{A}^e_{vw}\brc{\bra{n-1}\mu^2+\sigma^2}\brc{\bra{1-\delta_{0w}}\delta_{0v}}\nonumber\\&-\sum_{v,w}\vc{A}^e_{vw}\brc{\bra{n-1}\mu^2+\sigma^2}\brc{\bra{1-\delta_{0v}}\delta_{0w}}\nonumber\\&+\sum_{v,w}\vc{A}^e_{vw}\brc{\delta_{vw}\sigma^2+\mu^2}\bra{1-\delta_{0v}}\bra{1-\delta_{0w}}		    }
Ordering the terms for $\mu$ and $\sigma$ respectively we can acquire the coefficient matrices $\vc{U}$ and $\vc{V}$,
\g{ \mean{\LP p_e ^2} =    \sum_{v,w}\vc{A}^e_{vw}  \mu^2 \brc{ \bra{1-\delta_{0v}}\bra{1-\delta_{0w}}+\delta_{0v}\delta_{0w}\bra{n-1}^2\right. \nonumber\\ \left. -\bra{n-1}\bra{\delta_{0v}\bra{1-\delta_{0w}} +\delta_{0w}\bra{1-\delta_{0v}}} \vphantom{\bra{n-1}^2}}\nonumber\\ + \sum_{v,w}\vc{A}^e_{vw}\sigma^2\brc{ \bra{1-\delta_{0w}}\bra{1-\delta_{0v}}\delta_{vw}-\delta_{0v}\bra{1-\delta_{0w}} \right. \nonumber\\ \left. -\delta_{0w}\bra{1-\delta_{0v}}+\bra{n-1}\delta_{0v}\delta_{0w}} \\
=    \sum_{v,w}\vc{A}^e_{vw}  \mu^2 \brc{ 1+\delta_{0v}\delta_{0w}n^2 -n\bra{\delta_{0v}+\delta_{0w}}}\nonumber\\ 
+ \sum_{v,w}\vc{A}^e_{vw}\sigma^2\brc{\delta_{vw}+\bra{n+1+\delta_{vw}}\delta_{0v}\delta_{0w}\right. \nonumber\\ \left. -\bra{1+\delta_{vw}}\bra{\delta_{0v}+\delta_{0w}}  }		\label{eq:mean_sq} }
We further suggest to expand this ansatz by introducing additional sources which act as clones of the very first one, i.e.\ we will have $s_p=s_q$ using the indices $p,q$ for sources and $m,n$ for sinks. Then conditions \eqref{eq:source1}, \eqref{eq:source2},\eqref{eq:source3} will become for $a$ sources and $b$ sinks (with $a+b=n$),
\alg{\mean{\sum_p s_p} & =-\mean{\sum_{u}s_u}=-b\mu\rightarrow \mean{s_p}=-\frac{b}{a}\mu \label{eq:source4}\\  \mean{s_p s_q} & =\frac{1}{a^2}\mean{\sum_{u,v}s_us_v}=\frac{1}{a^2} \bra{b\sigma^2 + b^2\mu^2} \label{eq:source5}\\ \mean{s_ps_u} & =-\frac{1}{a}\mean{\sum_{v}s_vs_u} =-\frac{1}{a}\bra{b\mu^2-\sigma^2} \label{eq:source6}}	 
And hence we may calculate the mean squared pressure and its coefficient matrices respectively as,
\g{
		 \mean{\LP p_e ^2} =   \sum_{v,w}\vc{A}^e_{vw}  \mu^2 \brc{ \frac{b^2}{a^2}\sum_{p,q}\delta_{pv}\delta_{qw} \right. \nonumber\\ \left. - \frac{b}{a}\sum_{p,m}\bra{\delta_{pv}\delta_{mw} +\delta_{mv}\delta_{pw}}+\sum_{m,n}\delta_{mv}\delta_{nw}}\nonumber\\+ \sum_{v,w}\vc{A}^e_{vw}\sigma^2\brc{ \frac{b}{a^2}\sum_{p,q}\delta_{pv}\delta_{qw} \right. \nonumber\\ \left. - \frac{1}{a}\sum_{p,m}\bra{\delta_{pv}\delta_{mw}+\delta_{mv}\delta_{pw}} +\sum_{m,n}\delta_{mv}\delta_{nw}\delta_{mn} }\label{eq:mean_sq_multi}
		 }
		 
\section{Simulating coupled, adapting networks}\label{apx:simulations}
\begin{figure}[h]
  \subfloat[Network \# 1]{	\label{fig:bilayer_nullity_supplement_a}\hspace{-.75cm} \includegraphics[scale=1.]{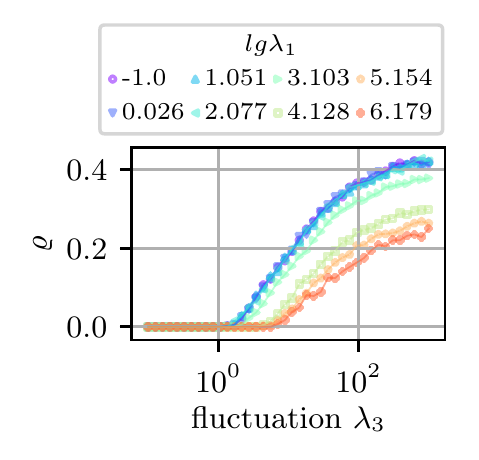}\hspace{-.35cm}
 \includegraphics[scale=1.]{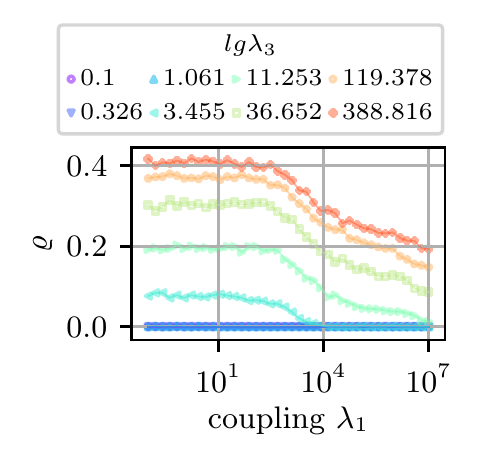}}\\
 \subfloat[Network \# 2]{	\hspace{-.75cm}\label{fig:bilayer_nullity_supplement_b} \includegraphics[scale=1.]{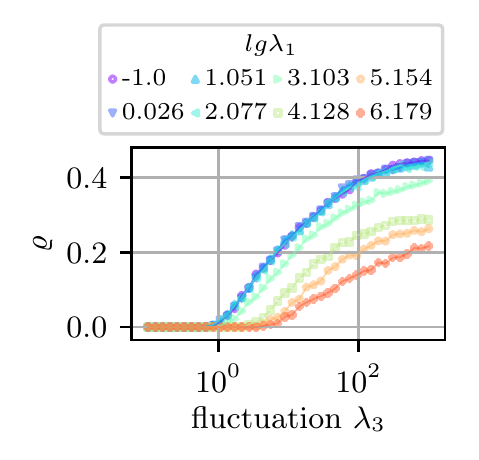}\hspace{-.35cm}
 \includegraphics[scale=1.]{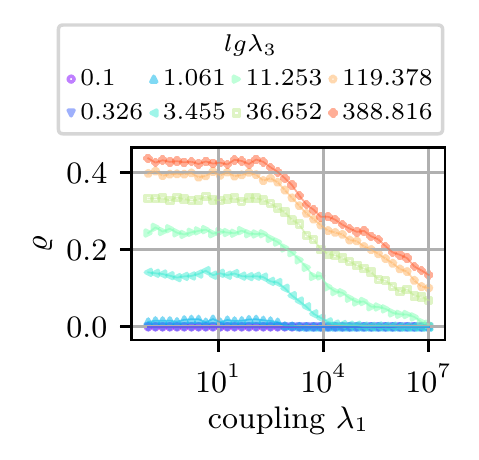}}\\
\caption{Nullity $\varrho$ transitions for repulsively coupled networks ($\varepsilon=-1)$: Symmetric fluctuation $\lambda_3$ induced nullity onset and coupling $\lambda_1$ induced nullity breakdown for the two networks  \subR{fig:bilayer_nullity_supplement_a}  and \subR{fig:bilayer_nullity_supplement_b} 
}\label{fig:bilayer_supplement_repulsive}
\end{figure}

\begin{figure}[h]
  \subfloat[Network \# 1]{	\hspace{-.6cm}\label{fig:bilayer_nullity_supplement_c} \includegraphics[scale=1.]{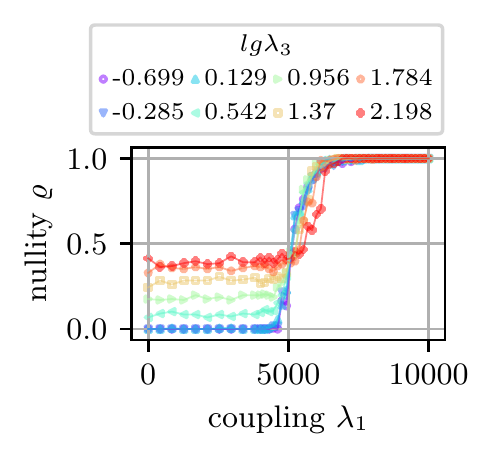}\hspace{-.35cm}
  \includegraphics[scale=1.]{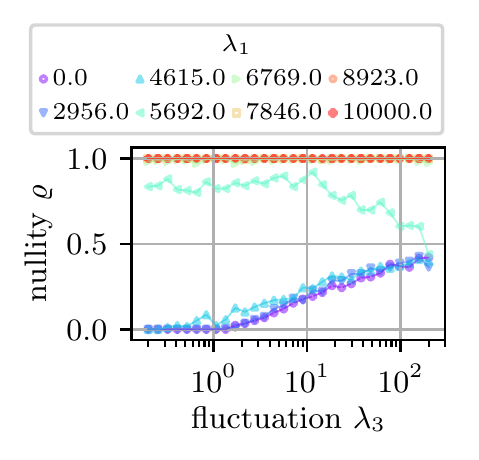}} \\\vspace{-0.35cm}
   \subfloat[Network \# 2]{\hspace{-.6cm}	\label{fig:bilayer_nullity_supplement_d} \includegraphics[scale=1.]{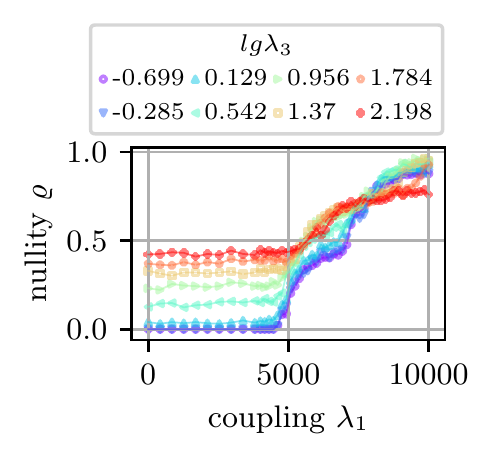}\hspace{-.35cm}
 \includegraphics[scale=1.]{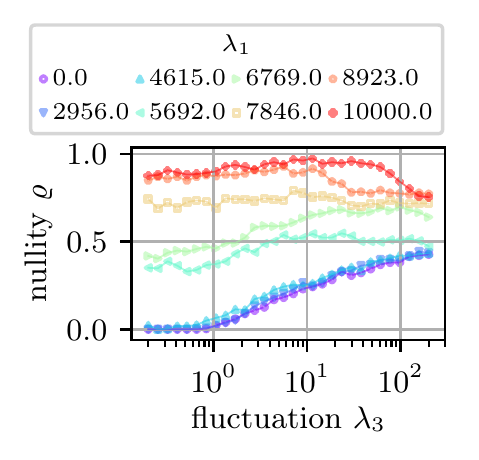}}
\caption{Nullity $\varrho$ transitions for attractively coupled networks ($\varepsilon=3)$: Slightly asymmetric coupling $\lambda_1$ induced nullity onset and fluctuation $\lambda_3$ induced nullity transition for the two networks  \subR{fig:bilayer_nullity_supplement_c}  and \subR{fig:bilayer_nullity_supplement_d} 
}\label{fig:bilayer_supplement_attractive}
		\end{figure}

\begin{figure}[h]
  \subfloat[Network \# 1]{  	\hspace{-.6cm}
\includegraphics[scale=1.]{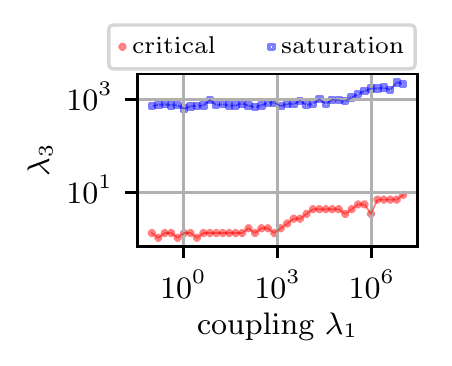}\label{fig:bilayer_onset_supplement_1a}}
\subfloat[Network \# 2]{	\hspace{-.5cm} \includegraphics[scale=1.]{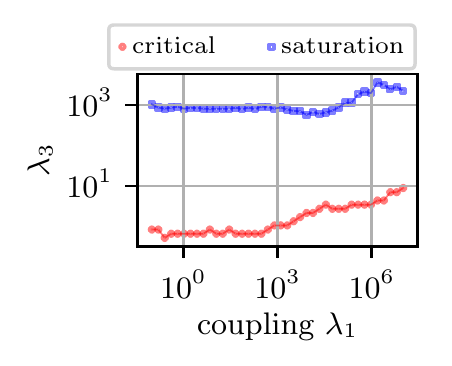}\label{fig:bilayer_onset_supplement_1b}}\\	
  \subfloat[Network \# 1]{	\hspace{-.6cm}	\includegraphics[scale=1.]{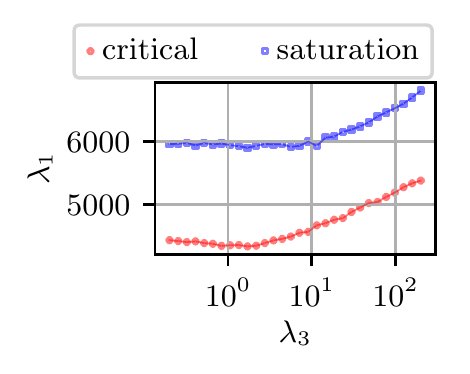}\label{fig:bilayer_onset_supplement_2a}}
  \subfloat[Network \# 2]{ 	\hspace{-.5cm} \includegraphics[scale=1.]{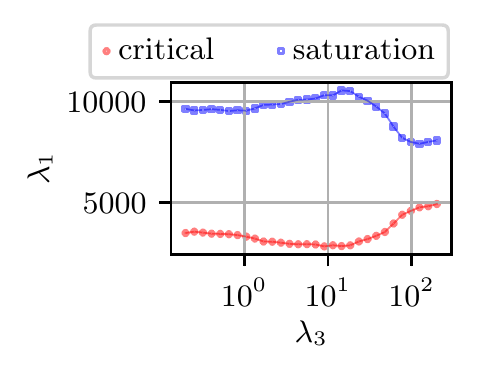}\label{fig:bilayer_onset_supplement_2b}}
 \caption{\subR{fig:bilayer_onset_supplement_1a}+\subR{fig:bilayer_onset_supplement_1b} Shift of onset and saturation points of the fluctuation $\lambda_3$ induced nullity transitions shift for repulsive coupling, here displayed for negative coupling exponent $\varepsilon=-1$.
\subR{fig:bilayer_onset_supplement_2a}+\subR{fig:bilayer_onset_supplement_2b}  Shift of onset and saturation points of the attractive coupling $\lambda_1$ induced nullity transitions shift for varying fluctuation $\lambda_3$, here displayed for positive coupling exponent $\varepsilon=3$
}
\end{figure}

In this section we show additional material of the simulations' results as performed in section \ref{sec:results} for both of the networks in comparison. In Figure \ref{fig:bilayer_supplement_repulsive} and \ref{fig:bilayer_supplement_attractive} we display the nullity transition curves for the cases of repulsive and attractive coupling. It may be noted here once again that all simulation parameters $\lambda_i$ were initialized identically for the two networks, meaning any deviations in $\varrho$ have to be caused by topological differences. This phenomenon is particularly apparent in the case of attractive coupling in cubic lattices where the nullity transition in the respective networks becomes smeared out differently for the two structures. Repeating the simulations for coupled networks consisting of complimentary Laves graphs or diamond lattices, we find this effect negligible. 
In the case of repulsive coupling  (Fig. \ref{fig:bilayer_onset_supplement_1a}, \ref{fig:bilayer_onset_supplement_1b}), both networks display a shift of the onset of fluctuation induced loops as well as a shift of the estimated saturation point. It may be noted though that a change of several orders of magnitude in the coupling $\lambda_1$ is necessary in order  to shift the fluctuation onset at all. It is nevertheless crucial to note that the overall nullity is reduced even in the saturated case which itself tends to be achieved only for significantly higher fluctuations $\lambda_3$. A similar trend of shifting may be observed in the case of attractive coupling  (Fig. \ref{fig:bilayer_onset_supplement_2a}, \ref{fig:bilayer_onset_supplement_2b}). The onset of full plexus recovery as well as the saturation onset develop on similar scales. Note that the trajectories for network \# 2 suggest the possibility of re-entrant behaviour between coupling and fluctuation dominated  nullity regimes.
\section{Scaling laws in coupled, noisy networks}\label{apx:scaling laws}
\begin{figure}[h]
			 \def\svgscale{0.38}
	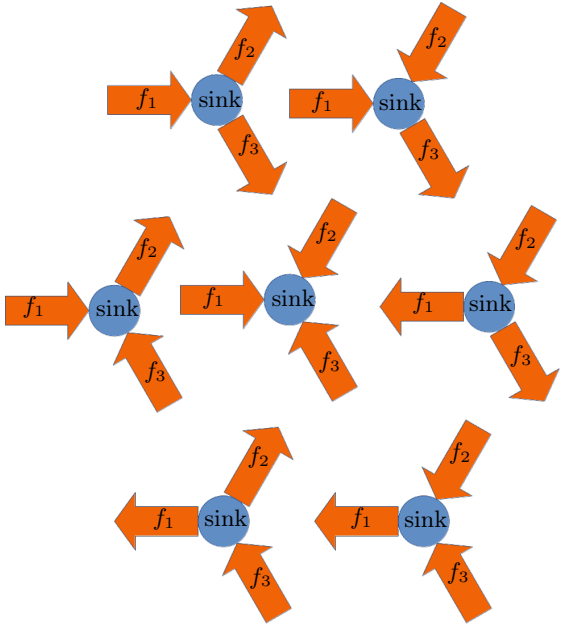
\caption{Scheme of identifying different flow combinations at a sink determining the respective sign of $\Theta_{ve}$, here shown for Y-branching point. Let's identify the sink as $i=0$ and and the respective flow carrying edges as $e\in \lbrace 1,2,3 \rbrace $ and define an incidence triplet $\vc{\Theta}=\bra{\Theta_{01},\Theta_{02},\Theta_{03}}$. Reading from left to right we have in  the top diagram:\\ $\vc{\Theta}=\bra{1,-1,-1}$, $\vc{\Theta}=\bra{1,1,-1}$\\ Middle diagram:\\ $\vc{\Theta}=\bra{1,-1,1}$, $\vc{\Theta}=\bra{1,1,1}$, $\vc{\Theta}=\bra{-1,1,-1}$\\ Bottom diagram:\\ $\vc{\Theta}=\bra{-1,-1,1}$, $\vc{\Theta}=\bra{-1,1,1}$}
\label{fig:floworder} 
		\end{figure}

\begin{figure}[h]
 \subfloat[]{\includegraphics[scale=.5]{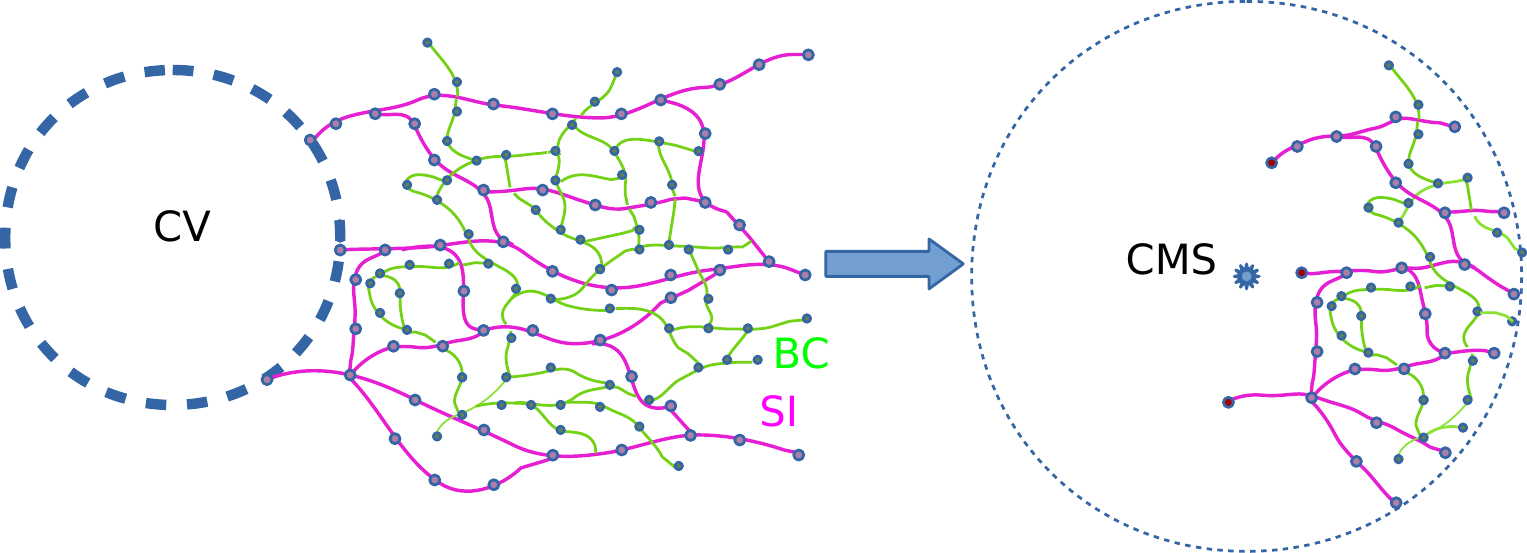}\label{fig:sketch_roi}}\\
  \def\svgscale{1.}
			\subfloat[]{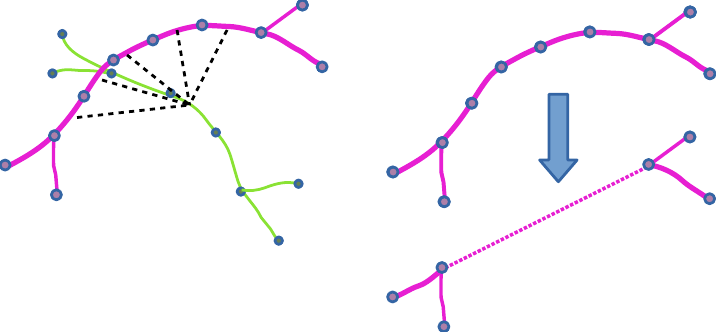\label{fig:sketch_coarse}}
	\caption{Setting the range of interest and coarse-graining the sinusoidal (SI) and bile canaliculi system (BC): \subR{fig:sketch_roi} Left: Vertices closest to the central vein (CV) are  identified as sinks. Right: Determine geometric center of mass of all sinks (CMS) and discard all components, vertices, edges outside a set perimeter.\subR{fig:sketch_coarse} Left: Deducting the distance of one path of a network to another path of the partner network by calculating the pair-wise distances of all path-segments and finding its minimum. Right: Coarse-graining paths into one effective edge, with new edge weight according to addition theorem $c^*_{eff}=\sum_e\frac{1}{c^*_e}$. }\label{fig:sketch}
			\end{figure}

\begin{figure*}[h]
\subfloat[]{\includegraphics[scale=1.]{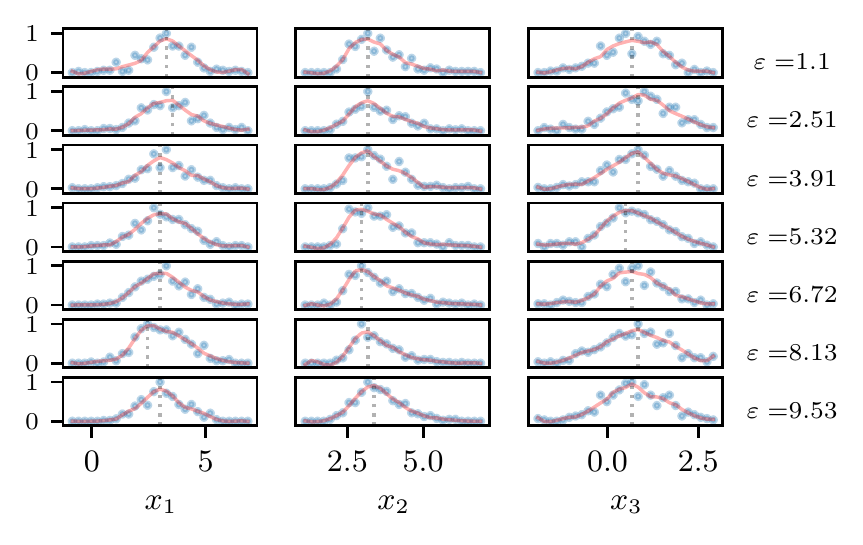}\label{fig:hist_attractive} }
\subfloat[]{\includegraphics[scale=1.]{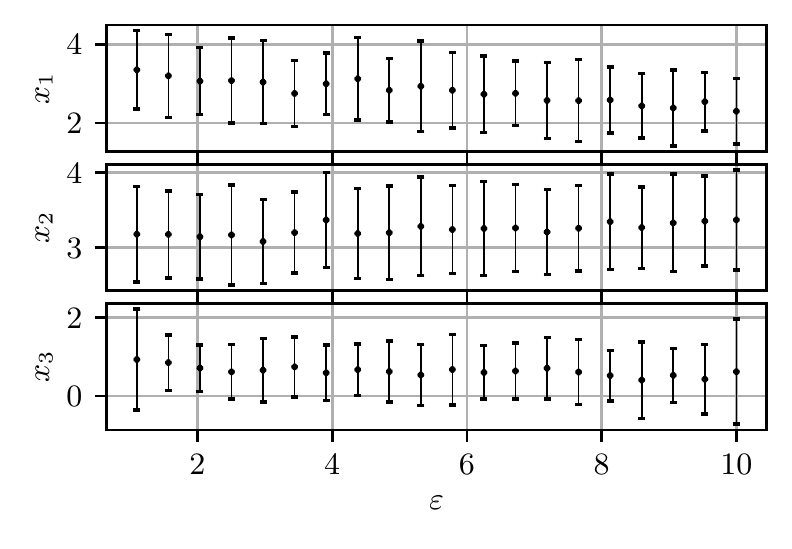}\label{fig:epsilon_attractive} }\\
\subfloat[]{\includegraphics[scale=1.]{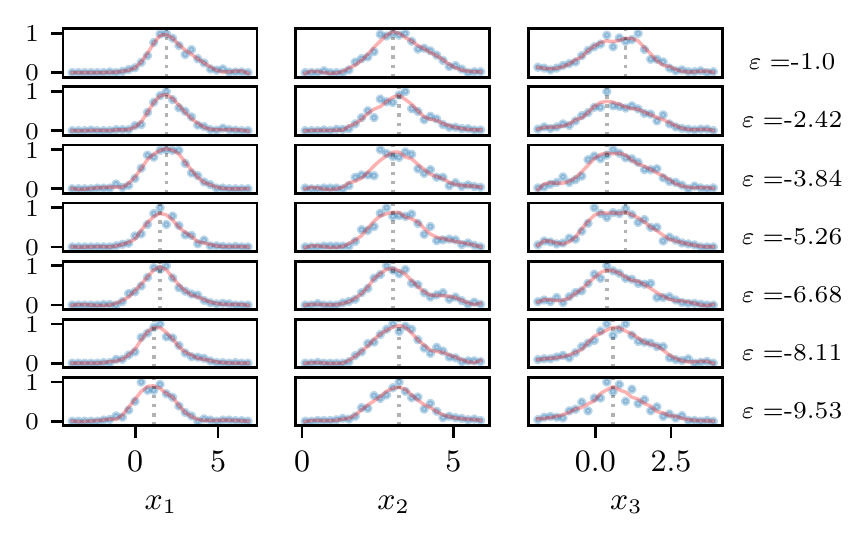}\label{fig:hist_repulsive} }
\subfloat[]{\includegraphics[scale=1.]{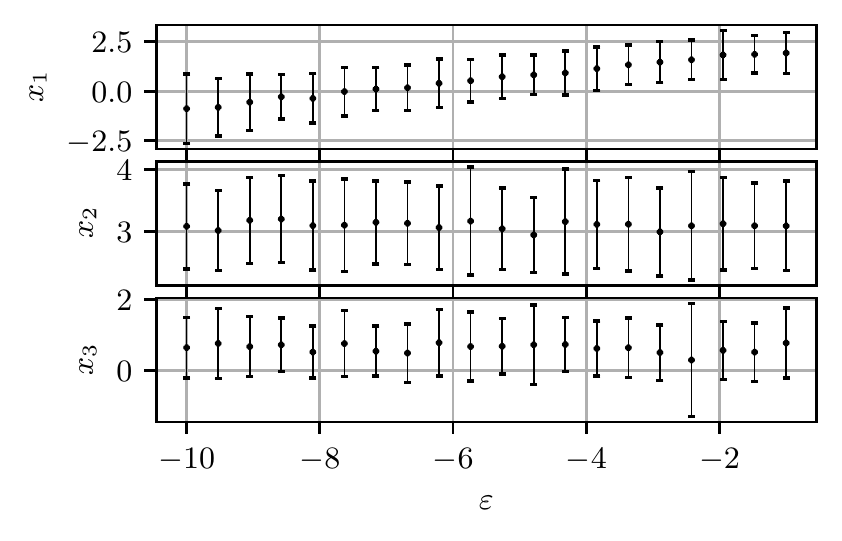}\label{fig:epsilon_repulsive} }
\caption{Estimated parameter distributions with rescaled axis $x_i=\log_{10}\lambda_i$ (in blue) for selected $\varepsilon$ in the sinusoidal system for $R=397\ \mu m$, $\delta= 30\ \mu m$. Applied a Savitzky-Golay regression (in red) to enable an effective peak identification. Subsequently, Gaussian fits are used to calculate the means and standard deviations of the smoothed distributions. \subR{fig:hist_attractive}+\subR{fig:epsilon_attractive}: Attractive Coupling \subR{fig:hist_repulsive}+\subR{fig:epsilon_repulsive}: Repulsive Coupling }\label{fig:hist}
\end{figure*}

This section is focused on  the geometric law discussed in section \ref{sec:estimation}. One can show that introducing fluctuations and coupling alter the classical form of Murray's Law $r_0^{3}=r_1^{3}+r_2^{3}$ in the following way: Given the Kirchhoff current law and rewriting it via Ohm's law, we get for all sink-vertices $i$,
		\alg{
			\sum_{e \in \inc(v)} f_e=s_v \Leftrightarrow &  \sum_{e \in \inc(v)} r_{e}^{*4} \LP p_{e}^*=s_v^*
			}
 Taking the average over all pressure configurations between two adaptation events  we get
  \alg{
   \sum_{e \in \inc(v)} r_{e}^{*4} \mean{\LP p_{e}^*}=\mean{s_v^*}
  \Leftrightarrow & \sum_{e \in \inc(v)} \Theta_{ve} r_{e}^{*4}\vert\mean{\LP p_{e}^*}\vert=\mean{s_v^*} \label{eq:current_law_dV}
		}
with an effective incidence factor $ \Theta_{ve}=\begin{cases}\pm 1\\ 0\end{cases} $ distinguishing between in- and outgoing flows on the relevant edges, see Figure \ref{fig:floworder}. In order to acquire the cubic form we substitute $r_e^*$ with the result of the stationary state's equations,
 \alg{ 
 \partd{t^*} r_e^* &=0\\ \Rightarrow 0 &= \bra{r_e^{*2}\mean{\LP p_e^{*2} }- \lambda_2} r_e^* + \lambda_1 \sgn(\varepsilon)\sum_{e'}F_{ee'}\LP r_{ee'}^{*,-\varepsilon}\\
\Leftrightarrow r_e^* & = \frac{\sqrt{\lambda_2-\lambda_1 r_e^{*-1}\sgn(\varepsilon)\sum_{e'}F_{ee'}\LP r_{ee'}^{*,\varepsilon-1}}}{\sqrt{\mean{\LP p_e^{*2}}}}\label{eq:stationary}
}
Plugging \eqref{eq:stationary} into \eqref{eq:current_law_dV} and rewriting ${\vert\mean{\LP p_{e}^*}\vert=\sqrt{ \mean{\LP p_{e}^*}^2 }}$, ${g^*_e\bra{\varepsilon}=\sgn(\varepsilon)\sum_{e'}F_{ee'}\LP r_{ee'}^{*,\varepsilon-1}}$ we get,
\alg{
	\sum_{e \in inc(v)} \Theta_{ve} \sqrt{\bra{\lambda_2-\lambda_1 r_e^{*-1}g^*_e\bra{\varepsilon}}\frac{\mean{\LP p_e^*}^{2}}{\mean{\LP p_e^{*2}}}}r_{e}^{*3}=\mean{s_v^*} \label{eq:current_law2}
}
We know from section \ref{apx:flow_fluctuation} that $\mean{\LP p_e^2}  = \sum_{vw}\vc{A}^e_{vw}\mean{s_v s_w}$ and may also deduce ${\mean{\LP p_e} ^2   =\sum_{vw}\vc{A}^e_{vw}\mean{s_v }\mean{s_w}}$. This enables us to calculate the ratio via the covariance,
\alg{
	\frac{\mean{\LP p_e^2}}{\mean{\LP p_e} ^2} & = \frac{\sum_{vw}\vc{A}^e_{vw}\mean{s_v s_w}}{\sum_{vw}\vc{A}^e_{vw}\mean{s_v }\mean{s_w}}\\ & =\frac{\sum_{vw}\vc{A}^e_{vw}\brc{\mean{s_v}\mean{s_w}+\mean{\bra{s_v-\mean{s_v}} \bra{s_w-\mean{s_w}}}}}{\sum_{vw}\vc{A}^e_{vw}\mean{s_v}\mean{ s_w}} \\ &=1+\frac{\sum_{vw}\vc{A}^e_{vw}\mean{\bra{s_v-\mean{s_v}} \bra{s_w-\mean{s_w}}}}{\sum_{vw}\vc{A}^e_{vw}\mean{s_v}\mean{ s_w}}
}
Substituting this into \eqref{eq:current_law2} and having $\frac{\mean{\LP p_e^2}}{\mean{\LP p_e}^2}=\frac{\mean{\LP p_e^{*2}}}{\mean{\LP p_e^*}^2}$, $H_e=\frac{\sum_{vw}\vc{A}^e_{vw}\mean{\bra{s_v-\mean{s_v}} \bra{s_w-\mean{s_w}}}}{\sum_{vw}\vc{A}^e_{vw}\mean{s_v}\mean{ s_w}}$ we get,
\alg{
	\sum_{e \in \inc(v)} a_e r_{e}^{*3}=\mean{s_v^*}\\ \text{ setting } a_e = \Theta_{ve} \sqrt{\frac{\lambda_2-\lambda_1 r_e^{*-1} g^*_e\bra{\varepsilon}}{1+H_e}} 
}
One may simplify this complex by considering the following: Assume that we have all randomly fluctuating sinks distributed uncorrelated yet identically as in section \ref{apx:flow_fluctuation}, \eqref{eq:mean_sq}, then we may set $\mean{s_v^*}=1$ and reevaluate $H_e$. This leads to the equation presented in section \ref{sec:estimation},
 \alg{\sum_{ \in \inc(v)} a_e r_e^{*3} & =1 \label{eq:murray}\\
 \text{ with } a_e & =\Theta_{ve}\sqrt{\frac{\lambda_2-\lambda_1 r_e^{*-1} g^*_e\bra{\varepsilon}}{1+\lambda_3 \frac{\delta\phi_e}{\phi_e}}}
}
 For experimental validation of \eqref{eq:murray} it will be necessary to know the networks vessel radii as well as  the sink/source pattern (although it may be sufficient to know where the system's source is and to consider every other node as sink). Considering Y-branchings of low sink/source value (points of negligible secretion/leakage) such that $| a_e r_e^{*,3}| \gg 1$,  further setting the index for the largest vessel to zero ( and accordingly increasing for the other vessel pieces at the branching)  may write,
\alg{
	a_0r^{*3}_0=a_1r^{*3}_1+a_2r^{*3}_2
}
In Figure \ref{fig:sketch} we show the applied reduction procedure as used in the section \ref{subsec:estimating_real}. 
First, the vertices in the sinusoidal network (SI) which are closest to the central vein (CV) are identified. Using these a geometric center of mass(CMS) is calculated and used as the center of a sphere of radius $R$, representing the range of interest (ROI). Any other components, vertices or edges of any network positioned ROI, see Figure \ref{fig:sketch_roi}. Next, all branching points in the sinusoidal network are identified and all paths $\vc{p}=\bra{e_i,...,e_j}$ consisting of edges $e_i$ which start from these points. We proceed for the canaliculi the same way and then check for each segment of a path $\vc{p}$ whether there is another segment of another network's path $\vc{p}'$ inside a perimeter $\delta$ . If so, these paths count as affiliated. Then we merge all edges along a path towards a single edge by using the conventional addition theorems for series of conductivities, as $c_{eff}=\sum_e\frac{1}{c_e}$, see Figure \ref{fig:sketch_coarse}. When confronted with datasets of real spatial transport network such as blood capillaries and secretion channels in \ref{subsec:estimating_real} it becomes important to note that the chosen ROI, namely the radius $R$ and the affiliation perimeter $\delta$ will affect the estimation of the parameters $x_i=\log\bra{\lambda_i}$.
\begin{figure}[h]
\includegraphics[scale=1.]{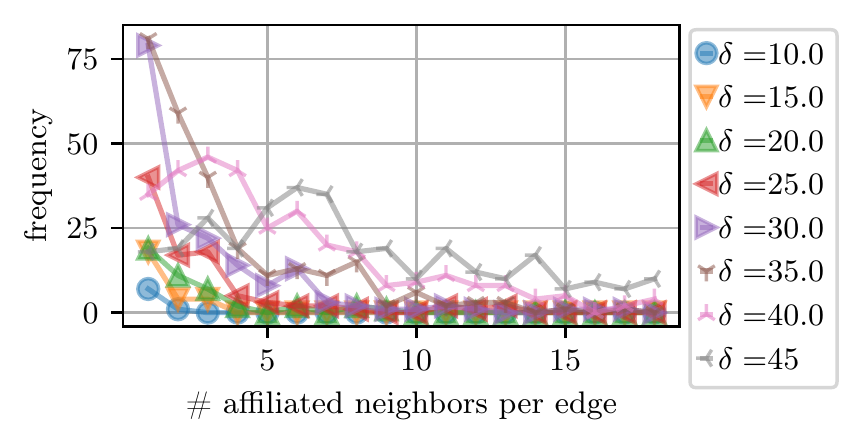}
\caption{ Histogram displaying the number of affiliated neighbors (edges) found for a given radius around a vessel (threshold, all values the inset in $\mu m$). Total number of pairings found overall for given threshold in the inset for set range of interest $R=397\ \mu m$.}\label{fig:hist_affiliation} 
\end{figure}
For example see Figure \ref{fig:hist_affiliation}, increasing the affiliation parameter $\delta$ will naturally increase the number of edges affiliated with each other, possibly linking vessel structures of several consecutive neighborhoods. \\
Further we attempt to screen through the range of potential coupling exponents $\varepsilon$ in order to find the $\lambda_i$ distributions as displayed in Figure \ref{fig:murray_par_fit}, section \ref{subsec:estimating_real}. In Figure \ref{fig:hist} we display the explicit distributions for selected $\varepsilon$ depicting the different coupling regimes. We smooth the datasets with a Savitzky-Golay filter (red fit line in \ref{fig:hist}) and identify the local maxima which display a prominence of at least $10 \%$ of the maximal function value. We subsequently use Gaussian fits at the identified peak's positions to acquire the mean values and standard deviations as depicted in \ref{fig:epsilon_attractive},\ref{fig:epsilon_repulsive}.


\end{document}